\pdfoutput=1
\documentclass[a4paper,UKenglish,cleveref,numberwithinsect,thm-restate]{lipics-v2021}
\hideLIPIcs

\usepackage{stmaryrd}
\usepackage[disable]{todonotes}
\usepackage{mathpartir}
\usepackage{appendix}

\crefformat{section}{\S\kern0.8pt#2#1#3}
\crefformat{subsection}{\S\kern0.8pt#2#1#3}

\DeclareSymbolFont{pxfontssymbolsC}{U}{pxsyc}{m}{n}
\DeclareMathSymbol{\Coloneqq}{\mathrel}{pxfontssymbolsC}{70}

\newcommand{\pcfv}{\ensuremath{\mathsf{PCF_v}}}

\newcommand{\ssp}{\ensuremath{\mathsf{SSP}}}

\newcommand{\sheaves}{\mathop{\mathsf{Sh}}}

\newcommand{\sets}{\ensuremath{\mathsf{Set}}}
\newcommand{\op}{\mathsf{op}}

\newcommand{\vsets}{\mathsf{vSet}}
\newcommand{\vseq}{\mathrm{V}}

\newcommand{\cterm}{\mathrel{\vdash^{\mathbf{c}}}}
\newcommand{\vterm}{\mathrel{\vdash^{\mathbf{v}}}}
\newcommand{\nat}{\mathsf{nat}}
\newcommand{\tzero}{\mathsf{0}}
\newcommand{\tone}{\mathsf{1}}
\newcommand{\rec}[3]{\mathsf{rec}\,#1\,#2.\,#3}
\newcommand{\letin}[3]{\mathsf{let}\, #1=#2\, \mathsf{in}\, #3}
\newcommand{\inl}[1]{\mathsf{inl}\,#1}
\newcommand{\inr}[1]{\mathsf{inr}\,#1}
\newcommand{\vzero}{\mathsf{zero}}
\newcommand{\suc}[1]{\mathsf{succ}(#1)}
\newcommand{\lbd}[2]{\lambda #1.\,#2}
\newcommand{\ret}[1]{\mathsf{return}\,#1}
\newcommand{\casesum}[5]{\mathsf{case}\, #1\, \mathsf{of}\, \{\mathsf{inl}\, #2 \rightarrow #3,\ \mathsf{inr}\, #4\rightarrow #5\}}
\newcommand{\casenat}[4]{\mathsf{case}\, #1\ \mathsf{of}\ \{\mathsf{zero}\rightarrow #2,\ \mathsf{succ}(#3)\rightarrow #4\} }
\newcommand{\caseempty}[1]{\mathsf{case}\, #1\, \mathsf{of}\ \{\}}

\newcommand{\den}[1]{\llbracket #1\rrbracket}

\newcommand{\Deltavsets}{\Delta_{\mathsf{vSet}}}
\newcommand{\liftvsets}{\mathop{L_\vsets}}
\newcommand{\odelta}{\mathcal O_\Delta}
\newcommand{\id}[1]{\mathsf{id}_{#1}}
\newcommand{\image}{\mathop\mathsf{im}}

\newcommand{\Deltag}{\Delta_{\mathcal G}}
\newcommand{\liftg}{\mathop{L_{\mathcal G}}}

\newcommand{\omegag}{\omega_{\mathcal G}}
\newcommand{\baromegag}{\bar\omega_{\mathcal G}}

\newcommand{\logval}[1]{\triangleleft^{\mathsf{val}}_{#1}}
\newcommand{\logcomp}[1]{\triangleleft^{\mathsf{comp}}_{#1}}

\newcommand{\partialmap}{\rightharpoonup}

\newcommand{\ssppar}{\ssp_\bot}

\newcommand{\incl}[1]{\mathop\mathsf{in}_{#1}}

\title{Recursion and Sequentiality in Categories of Sheaves}

\author{Cristina Matache}{University of Oxford, UK}{}{}{Research supported by an EPSRC studentship and Balliol College and Clarendon Fund scholarships.} 
\author{Sean Moss}{University of Oxford, UK}{}{}{Research supported by a Junior Research Fellowship at University College, Oxford.}
\author{Sam Staton}{University of Oxford, UK}{}{}{Research supported by a Royal Society University Research Fellowship and the ERC BLAST grant.}

\authorrunning{C. Matache, S. Moss and S. Staton} 

\Copyright{Cristina Matache, Sean Moss, and Sam Staton} 



\nolinenumbers 

\EventEditors{Naoki Kobayashi}
\EventNoEds{1}
\EventLongTitle{6th International Conference on Formal Structures for Computation and Deduction (FSCD 2021)}
\EventShortTitle{FSCD 2021}
\EventAcronym{FSCD}
\EventYear{2021}
\EventDate{July 17--24, 2021}
\EventLocation{Buenos Aires, Argentina (Virtual Conference)}
\EventLogo{}
\SeriesVolume{195}
\ArticleNo{25}

\begin{document}

\maketitle

\begin{abstract}
  We present a fully abstract model of a call-by-value language with higher-order functions, recursion and natural numbers, as an exponential ideal in a topos.  Our model is inspired by the fully abstract models of O'Hearn, Riecke and Sandholm, and Marz and Streicher. In contrast with semantics based on cpo's, we treat recursion as just one feature in a model built by combining a choice of modular components.
\end{abstract}

\section{Introduction}

This paper is about building denotational models of programming languages with recursion by using categories of sheaves. The naive idea of denotational semantics is to interpret every type $A$ as a set of values $\den A$, every typing context $\Gamma$ as a set of environments $\den\Gamma$, and every term $\Gamma\vdash t:A$ as a partial function $\den t:\den\Gamma\rightharpoonup\den A$, so that composing terms corresponds to composing functions. A more general approach says that a `denotational model' is a category with enough structure, such as a category of sets, so that we regard $\den\Gamma$ and $\den A$ as objects of that category, and $\den t$ as a morphism. In our work here, we work in various categories of sheaves, so that $\den\Gamma$ and $\den A$ are sheaves, which is not far from the naive set-theoretic idea because categories of sheaves are often regarded as models of intuitionistic set theory. As we will explain, each category of sheaves is captured by a small site, and by combining or comparing sites we can combine and compare different denotational models of programming languages.

We illustrate this by combining sites to give a fully abstract model of a call-by-value PCF.
Full abstraction means that two terms $t$, $u$ are interpreted as equal functions ($\den t=\den u$)  if and only if they are contextually equivalent. In PCF, which is a simple functional language, the main challenge for full abstraction is to capture the fact that PCF is sequential, in that it does not have any primitives for parallelism.

Our model is inspired by earlier models that were \emph{not} explicitly sheaf-theoretic~\cite{DBLP:journals/iandc/OHearnR95,DBLP:journals/iandc/RieckeS02a,DBLP:books/daglib/0018087}. Our fully abstract model is built by combining many different sites which include one for recursion and that happen to include sites that will turn out to give full definability with truncated natural numbers. Overall, this truncated full definability can be used to prove full abstraction of the model.

Although the focus of this paper is on a simple PCF-like language, a broader agenda is to combine this analysis of recursion and sequentiality with recent sheaf-based models for other phenomena, including concurrency~(e.g.~\cite{cellular-howe}), differentiable programming~\cite{sherman-michel-carbin,huot-staton-vakar}, probabilistic programming~\cite{qbs}, quantum programming~\cite{lmz-quantum} and homotopy type theory~\cite{cubical-sets}. The broader context, then, is to use sheaf-based constructions as a principled approach to building sophisticated models of increasingly elaborate languages. 

If the reader is familiar with synthetic domain theory, they may regard the contribution of this paper as an account of full abstraction in that tradition: at a high level we are merging the sheaf model of~\cite{fiore-rosolini-h} with the Kripke model of~\cite{DBLP:journals/iandc/OHearnR95}. We give a survey of SDT in \Cref{sec:sdt-review}.

\medskip

We now introduce the key ideas of our paper: to consider a general theory of `normal' models of PCF (\Cref{sec:intro-models}) and then to build a fully abstract one by combining certain sites (\Cref{sec:intro-fullabs}).

\subsection{Normal models of PCF}\label{sec:intro-models}

The key general definition of our paper is that of `normal model' (\Cref{def:model:fgcbv}). This has three components: a sheaf category; it has a well-behaved notion of partial function; and it supports recursion. 
We now discuss these three components. We motivate with the example of
the extended vertical natural numbers: the linear order ${\vseq=\{0\leq 1\leq \dots \leq n \leq \dots \infty\}}$. It is informally an interpretation of the ML datatype \verb|datatype v = succ of (unit -> v)|, or \verb|data V = Succ V| in Haskell, and it is widely regarded as a source of recursion (e.g.~\cite{crole-pitts}).

\begin{description}
  \item[Sheaf categories.]
We interpret types of the language as sheaves and terms as natural transformations between them. Following our motivating example, a \emph{(concrete) v-set} is a set~$X$ together with a given set $C_X\subseteq [\vseq\to X]$ of chains with endpoints; these should be closed under pre-composition with Scott-continuous functions of $\vseq$ and contain all constant functions.
For example, any cpo~$X$ can be regarded as a concrete v-set where the chains are the chains in $X$ with their limits. The concrete v-sets form the (concrete) sheaves on the one object category~$\mathbb{V}$ whose morphisms are Scott-continuous functions $\vseq\to \vseq$
(\Cref{sec:presheaves-vert-nat}).
It is helpful to bear in mind two views of this category, or any category of sheaves:
\begin{itemize}
\item  The external view is that the sheaves comprise sets with infinitary logical relations (of arity $\vseq$). The invariance property has the flavour of a Kripke structure, so they are similar to Kripke logical relations.
\item The internal view is that the category of sheaves is a model of intuitionistic set theory, with a special object $\vseq$ for which all functions $\vseq\to\vseq$ are continuous. 
\end{itemize}
\item[Partial functions with semidecidable domains.]
Our programming language contains functions that might not terminate, and so programs correspond to partial functions. Intuitively, we should only consider partial functions with a semidecidable domain. We formalize this by requiring that a normal model have a specified sheaf~$\Delta$ of `semidecidable truth values' (\Cref{sec:partial-maps}, \Cref{def:gen:semidec:sub}). For example, in concrete v-sets we pick $\Delta=\{0\leq 1\}$ with $C_\Delta\subseteq[\vseq\to \Delta]$ the characteristic functions of infinite or empty up-sets.
In general, a choice of object $\Delta$ induces a `lifting' monad~$L$. So we can program with partial functions $X\to L(Y)$ using Moggi's monadic metalanguage~\cite{moggi-metalanguage}.

\item[Recursion via orthogonality.]
Among the v-sets, there is a canonical sheaf $\vseq$, but actually we can construct an analogous sheaf~$\bar\omega$ in any sheaf category with a semidecidable truth object~$\Delta$,
by taking a limit of a chain (\Cref{sub-sec:vertical-natural-numbers}). We can also define a \emph{non}-extended vertical natural numbers sheaf~$\omega$ by taking a \emph{co}limit of a chain; in v-sets this is the set $\{0\leq 1\leq\dots\}$ without an endpoint, with chains all the eventually constant ones. 

Our language has recursion, and we interpret recursive definitions in a sheaf~$A$ by using Tarski's fixed point theorem, by building a chain and taking its formal limit. This can be done in a canonical way when~$A$ is \emph{complete}, which we define in terms of \emph{orthogonality}.
The conditions says that the morphism $A^{\bar\omega}\to A^\omega$ induced by $\omega\subseteq \bar\omega$ is an isomorphism: intuitively, every chain has a canonical upper bound (\Cref{sec:tarski}).
We give a recipe for showing that~$A$ is complete for the interpretation of any type (\Cref{sec:recursion-partial-maps}).
\end{description}

Recall that cpo's can be regarded as v-sets,.
The constructions of product, function cpo, and lifting are all preserved by the inclusion functor hence the interpretation in v-sets is equivalent to the usual one in cpo's.
The point is that we can now follow the same kind of interpretation in any sheaf category with this structure, and we can combine our site~$\mathbb{V}$ with other sites, as we now explain.

\subsection{Combining sites and full abstraction}
\newcommand{\Syn}{\mathcal{S}\kern-0.85pt\mathit{yn}}
\newcommand{\unit}{\mathsf{unit}}
\label{sec:intro-fullabs}
In \Cref{sec:kripke:rel}, we build a sheaf category that is a normal model for our variant of PCF, that we show to be fully abstract in \Cref{thm:full-abstraction}. Our argument is based on full definability: every morphism has a syntactic counterpart.

Our construction in \Cref{sec:fully-abstract-model} is non-syntactic, but by way of motivation we first consider a site built from the syntax of PCF. First, let us define a syntactic `semidecidable subset' of a type $\tau$ to be a definable function
$s:\tau\to\unit$, i.e.~it will either terminate or diverge. Now we temporarily define a category
$\Syn$ where the objects are pairs $(\tau,s)$ of a type $\tau$ and a semidecidable property. A morphism $f:(\tau,s)\to(\tau',s')$ is a definable function $f:\tau\to \tau'$
such that $s=\big(\lambda x.\,f(x);()\big)$ and $f=\big(\lambda x.\,s(x);\mathsf{let}\,y=f(x)\,\mathsf{in}\,s'(y);y\big)$. 
In other words, the morphisms of this category should be regarded as total maps on their given domains. 

The presheaf category $[\Syn^\op,\sets]$ nearly satisfies all the requirements of a normal model, and since the Yoneda embedding $\Syn\to[\Syn^\op,\sets]$ is always full and faithful, we almost have a model with full definability. There are two obstacles which we will explain how to bypass: the natural numbers are not preserved by the Yoneda embedding, and we would prefer a non-syntactic model. To resolve these issues we also need machinery for combining concrete sites.

\begin{description}
\item[Natural numbers objects and truncated definability.]
In a non-trivial sheaf category there are uncountably many morphisms $\mathbb{N}\to\mathbb{N}$. This is arguably a good thing, in that we can reason set-theoretically, but it means that we cannot have full definability because the syntax is countable. We follow Milner~\cite{milner-fully-abstract-models-of-typed-lambda-calculi} in considering, for each $n$, a version of PCF where any natural number $>n$ triggers divergence. For this truncated language, it is possible to impose a sheaf condition on the site~$\Syn$ so that the Yoneda embedding $\Syn\to \sheaves(\Syn)$ preserves the structure of the language.
Now, by combining sites for all possible $n$, together with $\mathbb V$ to include recursion, we end up with sufficient definability.
\item[Non-syntactic models.]
To avoid using the syntax of PCF in the definition of the model, we consider a broader semantic class of sites that we can show include ones with truncated full definability.
We assemble this broad class of sites by using a general method (\Cref{sec:ssp}) based on a semantic structure for sequentiality called `structural systems of partitions'~\cite{MARZ2000133,DBLP:books/daglib/0018087}.
\item[Combining sites and concreteness.]
PCF satisfies the context lemma, which is to say that the meaning of a term with free variables can be determined by substituting closed values for those variables. In a categorical semantics, since the terminal object interprets the empty context,
the context lemma indicates that we are working with categories $\mathcal E$ that are \emph{concrete} in the sense that the hom-functor $\mathcal E(1,-):\mathcal E\to \sets$ is faithful: in effect, we are working with a category of sets and functions.

Sheaf categories are not concrete in general. In fact, in future work we intend to use non-concrete sheaf categories to address non-well-pointed phenomena in semantics~\cite{levy-amb}. But to model PCF, we need to ensure that when we combine sites we preserve concreteness. To this end we introduce a notion of sum for concrete sites, and show that it is a way of building normal models~(\Cref{sec:sum-sites}). Moreover, as we show, there are structure preserving functors out of this sum (\Cref{prop:sum-monad-lifting}). 
\end{description}
In summary, we build our fully abstract model by taking the sum of all the concrete sites that can be built with structural systems of partitions, together with $\mathbb{V}$ for recursion. We then show that all the definable models arise, and hence obtain the definability property, from which we can deduce full abstraction.

\section{A categorical setting for recursion} \label{sec:gener-sett-recurs}
\newcommand{\bind}{\mathrel{\scalebox{0.5}[1]{$>>=$}}}
Recursion in a programming language is usually interpreted
using Tarski's fixed point theorem (e.g.~\cite[\S12.5]{huet-fscd}). Although this is usually phrased in terms of partial orders of some flavour, in this section we provide a general abstract categorical treatment (\Cref{thm:fixed-points-x}). 
We give a language and its interpretation in~\Cref{sec:high-order-lang}.

For this section we fix a cartesian closed category~$\mathbb C$ with a pointed strong monad~$L$.
Recall that a cartesian closed category allows us to interpret a terminating typed $\lambda$-calculus, and that a strong monad is a triple $(L,\{\eta_X:X\to L(X)\}_X,\{\bind_{X,Y}: L(Y)^X\to L(Y)^{L(X)}\})$ satisfying associativity and identity laws, which allows us to interpret impure computation.
A \emph{pointed} monad is one equipped with a natural family of maps $\bot_A : 1 \to L(A)$.
We will think of $L$ as a partiality monad, so that morphisms $\Gamma\to L(X)$ are thought of as programs that need not terminate.
Our main example is the category $\vsets$ with its lifting monad $\liftvsets$ given in \Cref{sec:presheaves-vert-nat}, and the category $\mathcal G$ with $\liftg$ given in \Cref{sec:fully-abstract-model} is another.
In the meantime, it might help the reader to think of the category whose objects are posets and whose morphisms are monotone maps which preserve all suprema of $\omega$-chains that exist, together with the monad that adds a new element to the bottom of a poset.
Then \Cref{def:complete:obj} below would pick out as a full subcategory the category of $\omega$-cpo's and $\omega$-continuous maps.

Many of the ideas in this section and in \cref{sec:partial-maps} are well established in synthetic/axiomatic domain theory. We review the literature in~\cref{sec:sdt-review}.
\subsection{Vertical natural numbers}\label{sub-sec:vertical-natural-numbers}
In this abstract setting, provided certain limits and colimits exist, we can construct objects analogous to the linear orders $(0\leq 1\leq 2\leq \dots)$ and
$(0\leq 1\leq 2\leq \dots\leq \infty)$, respectively called the \emph{finite} and \emph{extended vertical natural numbers}. The relationship between these is crucial for Tarski's fixed point theorem.  

We assume that the following sequential diagram has a limit $\bar\omega$:
\begin{equation}\label{eq:limit:diag}
  1 \xleftarrow{!} L1 \xleftarrow{L(!)} LL1 \xleftarrow{LL(!)} \ldots
\end{equation}
We think of this limit as the extended vertical natural numbers. In particular,
there is a morphism $\mathsf{succ}_{\bar\omega} : \bar\omega \to \bar\omega$ determined
by the cone over diagram~\eqref{eq:limit:diag} with apex $\bar\omega$ given by $\bar\omega \to L^n 1 \xrightarrow{\eta_{L^n 1}} L^{n+1} 1$ and $! : \bar\omega \to 1 = L^01$.
  There is another cone with apex $1$ given by  $1 \xrightarrow{\eta_{L^{n-1} 1} \circ \ldots \circ \eta_{1}} L^n 1$ which defines a morphism $\infty : 1 \to \bar\omega$.
Note that $\mathsf{succ}_{\bar\omega} \circ \infty = \infty$.

We also assume that the following diagram has a colimit $\omega$:
\begin{equation}\label{eq:colimit:diag}
  1 \xrightarrow{\bot_1} L1 \xrightarrow{L(\bot_1)} LL1 \xrightarrow{LL(\bot_1)} \ldots
\end{equation}
We think of this colimit as the finite vertical natural numbers. In particular,
there is a cocone over diagram~\eqref{eq:colimit:diag} with apex $\omega$ given by $L^n1 \xrightarrow{\eta_{L^n1}} L^{n+1}1 \to \omega$ which defines a morphism $\mathsf{succ}_\omega : \omega \to \omega$.
There is a canonical comparison map $i : \omega \to \bar\omega$ which comes from maps
$L^m1 \xrightarrow{L^m(\bot_1)} \ldots \xrightarrow{L^{n-1}(\bot_1)} L^n1$ for $m \leq n$ and $L^m \xrightarrow{L^{m-1}(!)} \ldots \xrightarrow{L^n(!)} L^n1$ for $m \geq n$.

  It is straightforward to check that $i \circ (\mathsf{succ}_\omega : \omega \to \omega) = (\mathsf{succ}_{\bar\omega} : \bar\omega \to \bar\omega) \circ i$.

  \subsection{Complete objects and fixed points}\label{sec:tarski}
  In the traditional poset-based setting, Tarski's fixed point theorem requires that every chain has a least upper bound. This completeness can be expressed in this abstract categorical setting because a morphism $\omega\to X$ can be thought of as a chain in~$X$.

Recall that an object $X$ is said to be \emph{right-orthogonal} to a morphism $f:A\rightarrow B$ if every map $A\rightarrow X$ factors uniquely through $f$. We can then make the following definition: 
  
  \begin{definition}\label{def:complete:obj}
    An object $X\in\mathbb C$ is \emph{$L$-complete} if it is right-orthogonal to the morphism $\mathsf{id}_A \times i : A \times \omega \to A \times \bar\omega$ for every $A \in \mathbb C$.
\end{definition}
For example, in the category of $\omega$-cpo's and continuous maps, all objects are complete for the usual lifting monad.
From~\Cref{sec:partial-maps} we will work in sheaf categories where one does not expect this.

The present abstract setting admits the following fixed point theorem. The theorem is about $L$-complete objects that are moreover $L$-algebras (i.e.~objects $X$ equipped with a morphism $L(X)\to X$ satisfying conditions).
In the poset setting, $L$-algebras are just partial orders with a least element. 
\begin{restatable}{theorem}{fixedpoints}\label{thm:fixed-points-x}
  Let $X \in \mathbb C$ be an $L$-algebra and $LX$ an $L$-complete object. Then for any map $g : \Gamma \times X \rightarrow X$ we can construct a fixed point $\phi_g: \Gamma \rightarrow X$ such that $\phi_g(\rho)=g(\rho,\phi_g(\rho))$.
\end{restatable}

Given an interpretation for a language in $\mathbb C$ such that types are $L$-complete objects, we can use \Cref{thm:fixed-points-x} to interpret fixed points suitable for call-by-value:
\begin{restatable}{corollary}{fpcorollary}\label{cor:cbv:recursion}
  Consider objects $\Gamma$, $A$, $B$ in $\mathbb C$ such that $L(LB^A)$ is a $L$-complete object. For a morphism $M: \Gamma \times LB^A \times A \rightarrow LB$ we can construct a fixed point $\mathsf{rec}_M:\Gamma\rightarrow LB^A$ such that: $\mathsf{rec}_M(\rho)(a)=M(\rho,\mathsf{rec}_M(\rho),a)$.
\end{restatable}
Both fixed points $\phi_g$ and $\mathsf{rec}_M$ are constructed in \Cref{app:fix}.

\section{Partial maps, semidecidability and recursion in toposes}
\label{sec:partial-maps}
In this section we keep fixed a Grothendieck topos $\mathcal E$.
(We will not assume deep familiarity with Grothendieck toposes, but we recall that they are cartesian closed categories with a particularly well behaved notion of subobject and also well-behaved limits/colimits; these toposes turn out to be exactly the categories of sheaves on sites, see~\S\ref{sec:sites-sheaves}.)
We suppose moreover that $\mathcal E$ comes with a suitable notion of `semidecidable subset', which is classified by an object~$\Delta$ of~$\mathcal E$ as follows.
\begin{definition}\label{def:gen:semidec:sub}

  For a fixed object~$\Delta$ and a fixed monomorphism $\top : 1 \rightarrowtail \Delta$, we say a subobject of $A$ is \emph{semidecidable} if it is a pullback of $\top$ along some map $A \to \Delta$.
  
  We say that $\top : 1 \rightarrowtail \Delta$ is a \emph{generic semidecidable subobject} if:
  \begin{itemize}
  \item for every semidecidable subobject $m: A' \rightarrowtail A$ there is precisely one map $\phi : A \to \Delta$ such that $m$ is the pullback of $\top$ along $\phi$;
  \item every $0 \rightarrowtail A$ is semidecidable;
  \item semidecidable monomorphisms are closed under composition.
  \end{itemize}
\end{definition}
Our notion is almost exactly what was called a `dominance' in \cite{rosolini-phd} and a `partial truth value object' in \cite{mulry-partial-map-classifiers-and-partial-cccs}.
The difference is our requirement that the empty subobjects be semidecidable.

Throughout this section we assume a fixed generic semidecidable subobject ${\top : 1 \rightarrowtail \Delta}$.
It is straightforward to show that semidecidable subobjects are closed under finite meets, including top subobjects, and stable under pullback.
Moreover, all coproduct inclusions are semidecidable.

A \emph{partial map} $A \partialmap B$ consists of a semidecidable subobject $A' \rightarrowtail A$ and a map $A' \to B$.
Partial maps form a category, which can be given directly or described as the Kleisli category for a certain strong monad $L_\Delta$, the \emph{lifting monad}.
The unit of this monad assigns to each object $B$ its \emph{partial map classifier} $B \rightarrowtail L_\Delta B$, which is characterized by the property that maps $A \to L_\Delta B$ correspond to partial maps $A \partialmap B$ (the domain of the partial map is given by pulling back the subobject $B \rightarrowtail L_\Delta B$).
It is well-known that this gives a strong monad on $\mathcal E$ \cite{mulry-partial-map-classifiers-and-partial-cccs,cockett-lack-restriction-categories-ii-partial-map-classification}, which is moreover commutative and an `equational lifting monad' in the sense of \cite{bucalo-fuhrmann-simpson}.
The fact that $0 \rightarrowtail 1$ is semidecidable means that $L_\Delta$ has a point $\bot_A : 1 \to L_\Delta A$.

\subsection{Recipes for complete objects}
\label{sec:recursion-partial-maps}

We now show that a large amount of recursion comes from the assumption of $L_\Delta$-completeness of the generic semidecidable $\Delta$.
Since we are working in a Grothendieck topos $\mathcal E$, the colimit $\omega_\Delta$ and limit $\bar\omega_\Delta$ arising from the lifting monad $L_\Delta$ exist and are preserved by products, as in \S\ref{sub-sec:vertical-natural-numbers}.
It is useful to consider a slight strengthening of the $L_\Delta$-completeness condition, which roughly says that an object is $L_\Delta$-complete with respect to partial maps.
\begin{definition}
  Let $\odelta$ be the class of maps in $\mathcal E$ which are pullbacks of maps $i \times \id A : \omega_\Delta \times A \to \bar\omega_\Delta \times A$ along semidecidable subobjects of $\bar\omega \times A$.
  Write \todo{SS: Better name? Predomain? Also, any intuition at all?}$\odelta^\boxslash$ for the class of objects right orthogonal to every map in $\odelta$.
\end{definition}
The following facts are standard and straightforward.
\begin{itemize}
\item $\odelta^\boxslash$ is contained in the class of $L_\Delta$-complete objects.
\item $\odelta$ is closed under the operations $(-)\times \id A$, under pullback along semidecidable subobjects, and under colimits in the arrow category of $\mathcal E$.
\item $\odelta^\boxslash$ is a reflective subcategory of $\mathcal E$, closed under limits, and an exponential ideal.
\end{itemize}
Every Grothendieck topos $\mathcal E$ admits a set $\mathcal S$ which generates $\mathcal E$ under colimits: if $\mathcal E$ is a presheaf topos, one may take $\mathcal S$ to be the representable presheaves; more generally if $\mathcal E$ is a sheaf topos take $\mathcal S$ to be the sheafiied representables.
Then it follows that the class $\odelta^\boxslash$ is equivalently the class of objects right orthogonal to a certain small subset of $\odelta$, those maps of the form $i \times \id A$ for $A \in \mathcal S$ taken from the generating set.

We summarize the following consequences of the assumption of $\Delta$ being $L_\Delta$-complete.
\begin{proposition}\label{prop:odeltaclass-omnibus-proposition}
  Suppose that $\Delta$ is $L_\Delta$-complete.
  \begin{itemize}
  \item $\Delta$ is in $\odelta^\boxslash$, and for $A \in \mathcal E$, $A \in \odelta^\boxslash$ iff $L_\Delta A$ is $L_\Delta$-complete iff $L_\Delta A \in \odelta^\boxslash$.
  \item $\odelta^\boxslash$ is closed under $L_\Delta$ and contains $0$.
  \item $\odelta^\boxslash$ is closed under $I$-indexed coproducts iff\/ $\sum_J 1 \in \odelta^\boxslash$ for some set $J$ with $|I| \leq |J|$.
  \end{itemize}
\end{proposition}
\begin{proof}[Proof notes]
  $\Delta$ being $L_\Delta$-complete means that there is a bijection between the semidecidable subobjects of $\omega_\Delta \times A$ and $\bar\omega_\Delta \times A$ for any $A$.
  From this, and the fact that $\Delta \cong L_\Delta 1$, one deduces the first claim.  
  Closure of $\odelta^\boxslash$ under $L_\Delta$ can be obtained directly, but also follows from Theorem 3.1 of \cite{dyckhoff-tholen-1987}, since $L_\Delta$ is a special case of a partial product functor.
  Finally, note that $\sum_J 1 \in \odelta^\boxslash$ implies that a $J$-indexed join of disjoint semidecidable subobjects of $A \times \bar\omega_\Delta$ is semidecidable iff the join of their pullbacks to $A \times \omega_\Delta$ is semidecidable.
\end{proof}

\section{A higher-order language with recursion}\label{sec:high-order-lang}

In this section we introduce the call-by-value calculus \pcfv\ whose models we will study in the rest of the paper. The calculus is an extension of the simply typed lambda calculus with binary products and sums and a type $\nat$ of natural numbers.
\pcfv\ is given as a fine-grained call-by-value calculus~\cite{levy-power-thielecke}, which means there is a syntactic distinction between values and computations. It includes a construct for defining recursive functions $(\rec{f}{x}{t})$ which should be thought of as the recursive definition of a function $f$, $f(x)=t$. There is also a construct for explicitly sequencing computations $\letin{x}{t}{t'}$.
\begin{align*}
\text{Types:}\quad  \tau &\Coloneqq \tzero \mid \tone \mid \nat \mid \tau + \tau \mid \tau \times \tau \mid \tau\rightarrow\tau\\
\text{Values:}\quad  v,w &\Coloneqq x \mid \star \mid \inl{v} \mid \inr{v} \mid (v,v) \mid \vzero \mid \suc{v} 
  \mid \lbd{x}{t} \mid \rec{f}{x}{t}\\
\text{Computations:}\quad  t  &\Coloneqq \ret{v} \mid  \casesum{v}{x}{t}{y}{t'} \mid \pi_1v \mid \pi_2v \mid v\ w
  \\
  &\mid \casenat{v}{t}{x}{t'}
  \mid  \letin{x}{t}{t'}
\end{align*}

There are two typing relations, one for values, $\vterm$, and one for computations, $\cterm$, defined as usual. We can define a big-step operational semantics in the usual way, by induction on types, as a relation $\Downarrow_\tau$ between a closed computation and a closed value, both of type $\tau$. The complete definitions appear in \Cref{app:typing:rules}. For example:
\begin{equation*}
  \inferrule{\Gamma,\,x:\tau \cterm t:\tau'}{\Gamma \vterm \lbd{x}{t}} \quad
  \inferrule{\Gamma,\, f:\tau\rightarrow\tau',\, x:\tau \cterm t:\tau'}{\Gamma \vterm \rec{f}{x}{t} : \tau\rightarrow \tau'} \quad
  \inferrule{t[(\rec{f}{x}{t})/f,\,v/x] \Downarrow_{\tau'} w}{(\rec{f}{x}{t})\ v \Downarrow_{\tau'} w}
\end{equation*}
The operational semantics gives the usual notion of contextual equivalence: two computations $t$ and $t'$ are contextually equivalent iff, for all contexts $C$ such that $C[t]$ and $C[t']$ are closed computations of ground type, $C[t] \Downarrow_\tau v \Leftrightarrow C[t']\Downarrow_\tau v$, and similarly for values.

\subsection{Denotational semantics}

We now outline the framework used for our denotational semantics of \pcfv.

\begin{definition}\label{def:model:fgcbv}
  A \emph{normal model} of \pcfv\ is a Grothendieck topos $\mathcal E$ together with a generic semidecidable subobject $1\rightarrowtail \Delta$ such that $L_\Delta(N_\mathcal{E})$ is a complete object for $L_\Delta$, where $N_\mathcal{E}=\sum_0^\infty1$.
\end{definition}

The interpretation of \pcfv\ types in any normal model $\mathcal E$ is given by
  $\den\tzero = 0$,  $\den\tone = 1$,  $\den\nat = {\textstyle \sum_0^\infty 1 = 1 + 1 + \ldots}$, 
  $\den{\tau\rightarrow \tau'}=\den{\tau}\Rightarrow L_\Delta\den{\tau'}$, $\den{\tau\times \tau'}=\den{\tau}\times \den{\tau'}$, and $\den{\tau +\tau'}=\den{\tau} + \den{\tau'}$.
  The interpretation for values and computations is standard. A value $\Gamma \vterm v:\tau$ is interpreted as a morphism $\den{\Gamma}\rightarrow \den{\tau}$ in $\mathcal E$. A computation $\Gamma \cterm t:\tau$ is a morphism $\den{\Gamma}\rightarrow L_\Delta\den{\tau}$. The term $(\rec{f}{x}{t})$ can be interpreted with the fixed point constructed in~\Cref{cor:cbv:recursion}.

Since $\Delta \cong L_\Delta 1$ (\Cref{def:gen:semidec:sub}) is a retract of $L_\Delta(N_{\mathcal E})$, the object $\Delta$ in a normal model is $L_\Delta$-complete.
Hence it follows from \Cref{prop:odeltaclass-omnibus-proposition} and its preceding discussion that all \pcfv\ types are interpreted as $L_\Delta$-complete objects in a normal model.

\section{Presheaves on the vertical natural numbers}
\label{sec:presheaves-vert-nat}

This section describes the category $\vsets$, an example of a normal model.
An object of $\vsets$, or a \emph{v-set}, is intuitively a set of points equipped with a abstract collection of limiting $\omega$-chains.
We ask that the chains be closed under the action of a monoid of reindexings.

Let $\mathbb V$ be the monoid of continuous monotone endomorphisms of the extended vertical natural numbers $\{0\leq 1\leq \dots \leq n\leq \dots \leq \infty\}$. 
As such, it is a one-object full subcategory of the category $\omega\mathsf{CPO}$ of $\omega$-cpo's.
Recall that the category $[\mathbb C^\op,\sets]$ of \emph{presheaves} on a small category $\mathbb C$ is the category with objects contravariant functors $F : \mathbb C^\op \to \sets$ and morphisms $F \to G$ natural transformations.

\begin{definition}\label{def:endo-vert}
$\vsets$ is the category $[\mathbb V^\op,\sets]$ of presheaves on $\mathbb V$.
\end{definition}
Equivalently, $\vsets$ is the category of sets equipped with an action of the monoid $\mathbb V$ with equivariant maps.
For $X \in \vsets$ we think of $X(\vseq)$ as a set of `abstract chains'.
We write $|X| = \vsets(1,X)$ for the set of global elements, thought of as `points'; note that we can also describe $|X|$ as the set of $x \in X(\vseq)$ such that $X(e)(x) = x$ for all $e \in \mathbb V(\vseq,\vseq)$.
Thus each abstract chain $s \in X(\vseq)$ gives an actual chain of points of $X$: $X(c_0)(s),X(c_1)(s),\ldots,X(c_\infty)(s)$, where $c_n : \vseq \to \vseq$ is the constant map with value $n$ for $n \in \mathbb N \sqcup \{\infty\}$.

The category $\omega\mathsf{CPO}$ embeds fully-faithfully into $\vsets$ by mapping an $\omega$-cpo $D$ to the set of $\omega$-chains in $D$ each equipped with their supremum.
V-sets in the image of $\omega\mathsf{CPO}$ have several special properties; one of them is that the map $X(\vseq) \to \sets(\mathbb N \sqcup \{\infty\},|X|)$ given by $s \mapsto \lambda n.X(c_n)(s)$ is injective.
An $X \in \vsets$ with this property is called a \emph{concrete v-set}, or \emph{concrete presheaf on $\mathbb V$} (we recall a generalization of this later in \Cref{def:concrete-sheaf}).
For a concrete v-set $X$, the abstract chains in $X(\vseq)$ may be identified with a set of functions $|\vseq| = \mathbb N \sqcup \{\infty\} \to |X|$ containing all constant functions and closed under precomposition with endomorphisms of $\vseq$.

The full embedding $\omega\mathsf{CPO} \hookrightarrow \vsets$ was already observed by Fiore and Rosolini \cite{fiore-rosolini-2sdt,fiore-rosolini-h}, who then considered a category of sheaves on $\mathbb V$ as a model of Synthetic Domain Theory.
Their sheaf condition is not relevant to our work here.
They consider a dominance in their sheaf category, which we treat as a generic semidecidable subobject in $\vsets$.
Let $\Deltavsets \in \vsets$ be the splitting in $\vsets$ of the idempotent $r_1 : \vseq \to \vseq$ given by $0 \mapsto 0$ and $x \mapsto 1$ for $x \geq 1$.
So $\Deltavsets(\vseq)$ can be identified with the set of monotone sequences $\mathbb N \to \{0,1\}$.
\begin{lemma}\label{lem:deltavsets-generic-semidecidable}
  $\Deltavsets$ is a generic semidecidable subobject, as in \Cref{def:gen:semidec:sub}. \end{lemma}
\begin{proof}[Proof notes]
  The most difficult part to check is that semidecidable monomorphisms are closed under composition.
  Given $\phi : A \to \Deltavsets$ classifying $m : B \rightarrowtail A$ and given $\psi : B \to \Deltavsets$, first note that $\psi$ admits an extension map $\psi' : A \to \Deltavsets$ where, for $x \in A(\vseq)$, $\psi'(x)$ is the greatest element of $\Deltavsets$ (in the lexicographic ordering) such that $\phi(\psi'(x)) = (1,1,\ldots)$ if it exists and $\psi'(x) = (0,0,\ldots)$ otherwise.
  Then if $\psi$ is the classifier of $n : C \rightarrowtail B$, the composite $mn : C \rightarrowtail A$ is classified by the map $\xi : A \to \Deltavsets$ where, for $x \in A(\vseq)$, $\xi(x)_i = \min\{\phi(x)_i,\psi'(x)_i\}$.
\end{proof}

Thus $\vsets$ admits a strong, pointed lifting monad $\liftvsets$, given by partial map classifiers as in the discussion following \Cref{def:gen:semidec:sub}.
This lifting monad can be explicitly given by~%
  ${(L_\vsets X)(\vseq) = \{ \bot \} +\ \sum_{n \in \mathbb N}(X(\vseq))_n}$
so it has a copy of the set $X(\vseq)$ for each $n\in \mathbb N$.
The action of an endomorphism $e$ on $\vseq$ is:
\begin{equation*}
  (L_\vsets X)(e)(s \in (X(\vseq))_n)
  =
    \begin{cases}
      \bot & \text{if } \image(e)\subseteq \{0,\ldots,n-1\} \\
      X(e')(s) \in (X(\vseq))_k & \text{if } e(\{0,\ldots,k-1\})\subseteq \{0,\ldots,n-1\},\\
      &e(k)>n-1,\ e'(i)=e(k+i)-n 
    \end{cases} \\
\end{equation*}
and $(L_\vsets X)(e)(\bot) = \bot$.
There is a ready intuition for $(\liftvsets X)(e)$ which is precise when $X$ is a concrete v-set: an element of $(X(\vseq))_n$ is a sequence $s$ of elements from $|X|$, to which we add $n$ $\bot$'s at the beginning.
The action $(L_\vsets X)(e)$ of an endomorphism $e$ of $\vseq$ is now just the standard reindexing of sequences by function composition $(\bot,\ldots,\bot,s)\circ e$.

We now show that $(\vsets,\Deltavsets)$ satisfies the conditions of a normal model (\Cref{def:model:fgcbv}) of \pcfv, which means showing that $\liftvsets(N_\vsets) = \liftvsets(\sum_0^\infty 1)$ is $\liftvsets$-complete.
It is straightforward to give the following explicit description of $\omega$ and $\bar\omega$:
for the $L_\vsets$ lifting monad on $\vsets$, the limit $\bar\omega$ is the representable $y(\vseq)$, and $i : \omega \to \bar\omega$ is the subobject of maps with bounded image (in particular, eventually constant).

\begin{lemma}
  $\Deltavsets$ is $\liftvsets$-complete.
\end{lemma}
\begin{proof}[Proof notes]
  Firstly, one checks that $\Deltavsets$ is orthogonal to $i : \omega \to \bar\omega$, since the maps into $\Deltavsets$ from $\omega$ or $\bar\omega$ are essentially just the eventually constant binary sequences.
  Then consider an extension problem $f : \omega \times A \to \Deltavsets$.
  Precomposing with the surjection on points $\coprod_{x \in |A|} \omega \times 1_{\{x\}} \to \omega \times A$, there is a unique extension to a map $\coprod_{x \in |A|} \bar\omega \times 1_{\{x\}}$.
  This gives a unique candidate extension of $f$ to $\bar\omega \times A$.
  To see that this is a valid morphism in $\vsets$, one simply checks that it maps $(\bar\omega\times A)(\vseq)$ into $\Deltavsets(\vseq)$.
\end{proof}

\begin{proposition}
  $\liftvsets(N_\vsets)=\liftvsets(\sum_0^\infty1)$ is $\liftvsets$-complete.
\end{proposition}
\begin{proof}[Proof notes]
  One observes that any map $\omega \to \liftvsets(\sum_0^\infty 1)$ or $\bar\omega \to \liftvsets(\sum_0^\infty 1)$ factors through one of the subobjects $\liftvsets(\iota_i) : \Deltavsets \cong \liftvsets 1 \rightarrowtail \liftvsets(\sum_0^\infty 1)$, where $\iota_i : 1 \to \sum_0^\infty 1$ is the $i$-th coproduct inclusion.
\end{proof}

Therefore, $(\vsets,\Delta_\vsets)$ is a normal model for \pcfv. Notice that $\den\tzero$ and $\den\tone$ are concrete v-sets. It is a standard fact that concrete presheaves are an exponential ideal, and that products and coproducts preserve concrete presheaves. Moreover, by straightforward inspection the lifting monad $L_\vsets$ preserves concreteness as well. Therefore, the \pcfv\ types are interpreted as \emph{concrete presheaves} in $\vsets$. This observation is useful for the proof of the next theorem (\Cref{app:adequacy}) because we only need to compare certain morphisms on their underlying points.

\begin{restatable}{theorem}{adequacy}\label{thm:adequacy-vsets}
  The pair $(\vsets,\Delta_\vsets)$ gives a sound and adequate model of \pcfv.
  \begin{itemize}
  \item Soundness: $t\Downarrow_\tau v \implies \den t = \eta_{\den\tau} \circ \den v \in \liftvsets\den\tau$.
  \item Adequacy: if $\tau$ is a ground type, $\den t = \eta_{\den\tau} \circ \den v \implies t \Downarrow_\tau v$. 
  \end{itemize}
\end{restatable}

\section{Sheaf conditions for sequentiality}\label{sec:kripke:rel}

In the previous section we used a simple index category, $\mathbb V$, to cut down the interpretation of \pcfv-types in $\sets$ to a model with recursion.
In this section we discuss the other index categories and their combinations, which we need for a fully abstract model. The motivation for the new index categories is that they each encapsulate a `prediction' of the underlying sets of the interpretations of types and the definable morphisms between them.
Roughly speaking, the relations force each prediction to arise as a full subcategory, including what turns out to be the correct prediction.

\subsection{Sites and sheaves}
\label{sec:sites-sheaves}
As the fully abstract model of \Cref{sec:fully-abstract-model} is given as the topos of sheaves on a site, we recall here some necessary definitions.
The standard reference is \cite{John02}, but for us all sites will be small.

A \emph{site} is a small category $\mathbb C$ equipped with a coverage $J$, where a \emph{coverage} $J$ on $\mathbb C$ is a set of covering families $(a, \{f_i \mid i \in I\})$ where $a \in \mathbb C$ and each $f_i$ is a morphism $f_i : a_i \to a$ with codomain $a$ such that, whenever $(a, \{f_i : a_i \to a \mid i \in I\}) \in J$ and $g : b \to a$ is in $\mathbb C$, there exists $(b, \{h_i : b_i \to b \mid  i \in I'\}) \in J$ such that, for all $i \in I'$, there exists $j \in I$ and $k : b_i \to a_j$ such that $f_j \circ k = g \circ h_i$.

Given a covering family $(a, \{f_i : a_i \to a \mid i \in I\}) \in J$ and a presheaf $F : \mathbb C^\op \to \sets$, a \emph{matching family} is a collection $(s_i \in F(a_i) \mid i \in I)$ such that for all $i,j \in I$, $b \in \mathbb C$, $g : b \to a_i$, and $h : b \to a_j$ we have $F(g)(s_i) = F(h)(s_j)$.
A \emph{sheaf} on the site $(\mathbb C,J)$ is a presheaf $F : \mathbb C^\op \to \sets$ such that for every covering family $(a, \{f_i : a_i \to a \mid i \in I\}) \in J$ and matching family $(s_i \in F(a_i) \mid i \in I)$ there is a unique element $s \in F(a)$ such that $F(f_i)(s) = s_i$ for all $i \in I$.
The element $s$ is called the \emph{amalgamation} of the matching family $(s_i)$.
The category of sheaves is denoted $\sheaves(\mathbb C,J)$.

The notion of coverage we have given here is a minimal one (see A2.1.9 of \cite{John02}).
There can be several coverages on one category $\mathbb C$ giving rise to the same collection of sheaves.
It is common to add saturation conditions to the coverage $J$ to tighten the correspondence between coverages and collections of sheaves, and also to assist calculation.
The following two are useful for us.
\begin{description}
\item [(M)] $J$ contains $(a, \{1_a : a \to a\})$ for all $a \in \mathbb C$.
\item [(L)] If $(a, \{f_i : a_i \to a \mid i \in I\}) \in J$ and $(b_i, \{g_{ij} : b_{ij} \to a_i \mid j \in J_i\}) \in J$ for $i \in I$ then $(a,\{f_ig_{ij} : b_{ij} \to a \mid i \in I, j \in J_i\}) \in J$.
\end{description}

\begin{example}
  Every small category $\mathbb C$ admits a `trivial' coverage, where $J = \emptyset$ and for which all presheaves on $\mathbb C$ are $J$-sheaves.
  For us, \emph{the trivial coverage} on $\mathbb C$ is given by $J = \{(a, \{1_a : a \to a\}) \mid a \in \mathbb C\}$, which has the same sheaves (all presheaves) but also satisfies (M) and (L).
\end{example}

A fundamental fact about $\sheaves(\mathbb C,J)$ is that it is a reflective subcategory of $[\mathbb C^\op,\sets]$, i.e.\ the inclusion functor $\sheaves(\mathbb C,J) \hookrightarrow [\mathbb C^\op,\sets]$ is full, faithful and possesses a left adjoint, which is called \emph{sheafification}.
A coverage is \emph{subcanonical} if all of the representable presheaves $\mathbb C(-,a)$ for $a \in \mathbb C$ are sheaves --- this means that sheafification leaves representables unchanged as functors $\mathbb C^\op \to \sets$.
The trivial coverage is subcanonical, but many useful coverages are not, and in this latter case the sheafified representables play a role analogous to that of the representable presheaves.
Hence we will sometimes find it useful to write $y$ for the composite $\mathbb C \to [\mathbb C^\op,\sets] \to \sheaves(\mathbb C,J)$ of the Yoneda embedding with sheafification.

\subsection{Concrete sites}
\label{sub-sec:concrete-sites}

We restrict our attention to a class of sites that are particularly convenient to work with.
Unlike the saturation conditions (M) and (L), these restrictions on $\mathbb C$ and $J$ do constrain the possible categories of sheaves.
Recall the following from \cite{dubuc-concrete-quasitopoi}.
\begin{definition}
  A \emph{concrete site} is a site $(\mathbb C,J)$ with a terminal object\/ $\star$ such that the maps $\mathbb C(a,b) \to \sets(\mathbb C(\star,a),\mathbb C(\star,b))$ are all injective, and $\coprod_{i \in I} \mathbb C(\star,a_i) \to \mathbb C(\star,a)$ is surjective for every covering family
 $(a,\{f_i : a_i \to a \mid i \in I \}) \in J$.
\end{definition}
In a concrete site it is convenient to define $|c| = \mathbb C(\star,c)$ for $c \in \mathbb C$ and to identify each morphism $c \to d$ with the induced function $|c| \to |d|$.
Thus $|-|$ is a faithful (but not necessarily full) functor $\mathbb C \to \sets$.
For a presheaf $X : \mathbb C^\op \to \sets$, we also write $|X| = X(\star) \cong \mathsf{Nat}(1,X)$.

A concrete site need not be subcanonical, but we can describe the sheafified representables as follows.
For any set $A$, the presheaf $\sets(|-|,A) : \mathbb C^\op \to \sets$ is a $J$-sheaf.
Every representable $\mathbb C(-,c)$ embeds into the sheaf $\sets(|-|,|c|)$ by concreteness and it follows that the sheafification $y(c)$ is the smallest subfunctor of $\sets(|-|,|c|)$ containing $\mathbb C(-,c)$ and closed under amalgamation.
When $J$ satisfies (M) and (L), then $y(c)$ is obtained by closing $\mathbb C(-,c)$ under amalgamation in just one step.

\begin{example}
  The category $\mathbb V$ as given \Cref{def:endo-vert} is not quite a concrete site, since it lacks a terminal object.
  However, as is well-known, the idempotent splitting of any small category has an equivalent presheaf category (see A1.1.19 of \cite{John02}).
  As the idempotent splitting of $\mathbb V$ contains a terminal object we are free to add it to $\mathbb V$, which we now treat as a concrete site with the trivial coverage.
\end{example}

A concrete site $(\mathbb C,J)$ is in particular a site, so it has a category $\sheaves(\mathbb C,J)$ of sheaves.
However, in this setting there is an especially useful subcategory.
\begin{definition}\label{def:concrete-sheaf}
  Let $(\mathbb C,J)$ be a concrete site.
  A \emph{concrete presheaf} is a presheaf $F : \mathbb C^\op \to \sets$ such that, for every $a \in \mathbb C$, the map $(F(x : \star \to a))_{x \in |a|} : F(a) \to \prod_{x \in |a|} |F|$ is injective.
  A \emph{concrete sheaf} is a concrete presheaf which is also a $J$-sheaf.
\end{definition}
The advantage of working with concrete presheaves is that if $Y$ is a concrete presheaf, and $X$ is any presheaf, then natural transformations $\alpha : X \to Y$ are determined by the function $\alpha_\star : |X| \to |Y|$.
As $Y(a) \subseteq \sets(|a|,|Y|)$, we can think of $Y$ as \emph{being} the set $|Y|$ together with an $\mathsf{ob}(\mathbb C)$-indexed family of relations.

We remark that concrete sheaves form a reflective subcategory, and so are closed under limits, and an exponential ideal.
All representables are concrete presheaves and concrete presheaves are closed under coproducts.
Since every concrete presheaf $X$ embeds into the concrete sheaf $\mathsf{Set}(|-|,|X|)$, every concrete presheaf injects into its sheafification and it follows that the concrete sheaves are closed under coproducts in sheaves.

\subsection{Defining concrete sites via systems of partitions}
\label{sec:ssp}
To help us define sites that we need for full abstraction, we first recall the category \ssp\ of Marz and Streicher~\cite{Mar00,MARZ2000133,DBLP:books/daglib/0018087}.
\begin{definition}
  Given a finite set $w$, a system of partitions $S^w$ is a set containing \emph{sets of disjoint subsets of $w$}, that is, (partial) partitions of $w$, and satisfying the following axioms:
  \begin{enumerate}
\item \label{ssp_1} $\{w\},\emptyset \in S^w$.

\item \label{ssp_2} (Refinement) $P,\,Q \in S^w$ and $U\in P$ imply that: $(P \setminus \{U\}) \cup (\{ U\cap V \mid V\in Q\} \setminus \{\emptyset\}) \in S^w$.

\item \label{ssp_3} $U,V \in P \in S^w$ implies that $(P \setminus \{U,V\}) \cup \{U\cup V\} \in S^w$.
\end{enumerate}
\end{definition}

The category \ssp\ has objects pairs $(w,\,S^w)$ of a finite set $w$ and a system of partitions $S^w$ for it. A morphism $f:(v,\,S^v)\rightarrow (w,\,S^w)$ is a set function $f:v\rightarrow w$ such that if $P=\{w_1,\ldots,w_n\} \in S^w$, then 
 $\{f^{-1}(w_1), \ldots, f^{-1}(w_n)\}\backslash\{\emptyset\} \in S^v$.
Composition is given by composition of functions.

The objects of \ssp\ encode the idea of a finite type $w$ together with a system of computable partial functions $w \partialmap \mathbb N$.
It may be helpful to think of these as potentially destructive measurements or observations on an unknown value of type $w$.
A partial partition $P \in S^w$ stands for an equivalence class of such functions which are undefined on $w \backslash \bigcup P$, constant on each $U \in P$, and which take distinct values on the members of distinct partition classes $U,V \in P$, where the equivalence is modulo a permutation of $\mathbb N$.
Axioms~\ref{ssp_1} and~\ref{ssp_3} correspond to such functions being closed under post-composition with all partial functions $\mathbb N \partialmap \mathbb N$, and containing all constant functions (including the totally undefined one).
Axiom~\ref{ssp_2} says that two computable functions $w \partialmap \mathbb N$ can themselves be sequenced together, say by checking whether $w \cterm t_1 : \nat$ returns $0$ and if so returning the outcome of $w \cterm t_2 : \nat$.

In light of the above, there is a natural notion of `semidecidable subobject' in \ssp.
For $P \in S^w$, there is a monomorphism $(\bigcup P, S^w\!\!\restriction_P) \rightarrow (w,S^w)$, where ${S^w\!\!\restriction_P} = \{Q \in S^w : \bigcup Q \subseteq \bigcup P\}$.
We say that any monomorphism isomorphic to one of this form is \emph{semidecidable}.
The corresponding notion of partial map admits partial map classifiers and hence a lifting monad.
This lifting monad is given by $L_\ssp(w,S^w)$ having underlying set $w \sqcup \{\bot\}$ and $S^{L_\ssp(w,S^w)} = S^w \sqcup \{\{w \sqcup \{\bot\}\}\}$.
We write $\ssppar$ for the Kleisli category of $L_\ssp$, or equivalently the category of partial maps in \ssp.

We are interested in faithful functors $F : \mathcal C \to \ssppar$.
The idea is that $\mathcal C$ stands for a system of finite types and definable partial functions between them, while $F$ equips each finite type with a system of measurements which is compatible with the partial functions in $\mathcal C$.
We now construct a topos $\mathcal E$ with generic semidecidable subobject such that $\mathcal E$ contains $\mathcal C$ as a full subcategory, and in which the observations on the \ssp-object $F(c)$ are precisely the partial maps $c \partialmap N_{\mathcal E} = \sum_0^\infty 1$ in $\mathcal E$.
\begin{definition}\label{def:icf}
  For a faithful functor $F : \mathcal C \to \ssppar$ the category $\mathcal I_{\mathcal C,F}$ is as follows.
  \begin{itemize}
  \item Objects: pairs $(c,U)$ where $c \in \mathcal C$ and $U = \bigcup P$ for some $P \in S^{F(c)}$ (equivalently $U=\emptyset$ or $\{U\} \in S^{F(c)}$ by axiom~\ref{ssp_3} of \ssp); and a distinguished terminal object $\star$.
  \item Morphisms $X \to Y$ are certain functions $|X| \to |Y|$, where $|(c,U)| = U$ and $|\star| = \{*\}$.
    When $X = (c,U)$ and $Y = (d,V)$, we take those functions $f : U \to V$ either constant or for which there is $\phi : c \to d$ in $\mathcal C$ such that $F(\phi) : F(c) \to L_\ssp(F(d))$ has domain $U$ and $F(\phi)(U) \subseteq V$.
    When either of $X,Y$ is $\star$, take all functions.
  \end{itemize}
\end{definition}
The category $\mathcal I_{\mathcal C,F}$ serves as `totalization' of $F : \mathcal C \to \ssppar$, by adding enough subobjects that every partial map can be represented by a total one.
It is not enough to take presheaves on $\mathcal I_{\mathcal C,F}$.
We need a coverage in order to force the coproduct $\sum_0^\infty 1$ to have the correct elements.
We emphasize that this is not merely an artefact arising from the sum types in \pcfv, it is necessitated by a normal model having $\nat$ interpreted as the coproduct $\sum_0^\infty 1$.
\begin{definition}
  Given a faithful functor $F : \mathcal C \to \ssppar$, the coverage $\mathcal J_{\mathcal C,F}$ has as covers families of partial identity maps $\{(c,U_i) \to (c,U)\}_{1 \leq i \leq n}$ where $P = \{U_1,\ldots,U_n\} \in S^{F(c)}$ and $\bigcup U_i = U$; and $\star$ is covered only by the identity.
\end{definition}
The following proposition is straightforward.
The main point to note is that axiom~\ref{ssp_2} of \ssp\ is required for the basic coverage axiom.
That same axiom is what gives us (L).
\begin{proposition}
  $(\mathcal I_{\mathcal C,F},\mathcal J_{\mathcal C,F})$ is a concrete site, satisfying the (M) and (L) axioms.
\end{proposition}

In $\sheaves(\mathcal I_{\mathcal C,F},J_{\mathcal C,F})$ we define $\Delta_{\mathcal C,F}$ where $\Delta_{\mathcal C,F}(c,U)$ is the set of subsets $U' \subseteq U$ where $(c,U')$ is an object of $\mathcal I_{\mathcal C,F}$, and $\Delta_{\mathcal C,F}(\star) = \{\emptyset,|\star|\}$.
The following is straightforward.
\begin{proposition}
  $\Delta_{\mathcal C,F}$ is a concrete sheaf and a generic semidecidable subobject in $\sheaves(\mathcal I_{\mathcal C,F},J_{\mathcal C,F})$.
  The lifting monad $L_{\mathcal C,F}$ preserves concrete sheaves.
\end{proposition}
We have the following explicit description of the lifting monad: $(L_{\mathcal C,F}A)(\star) = A(\star)+\{\bot\}$ and $(L_{\mathcal C,F}A)(c,U) = \coprod_{U' \subseteq U} A(c,U')$, where the coproduct is taken over all $U' \subseteq U$ such that there exists a partition $P\in S^{F(c)}$ such that $\bigcup P = U'$ (i.e.\ $(c,U')$ is an object of $\mathcal{I}_{\mathcal C,F}$).

One should not expect $\Delta_{\mathcal C,F}$ to be $L_{\mathcal C,F}$-complete.
It is only by summing with $\mathbb V$ as in the next section that we obtain a complete generic semidecidable subobject.
For this purpose it is still useful to describe the objects $\omega_{\mathcal C,F}$ and $\bar\omega_{\mathcal C,F}$ explicitly.
They are both concrete sheaves, and we can make the identifications $\bar\omega_{\mathcal C,F}(\star) \cong \mathbb N \sqcup \{\infty\}$ and $\omega_{\mathcal C,F}(\star) \cong \mathbb N$.
More generally, for $(c,U) \in \mathcal I_{\mathcal C,F}$, elements of $\omega_{\mathcal C,F}(c,U)$ are $\mathbb N$-indexed descending sequences of semidecidable subsets of $U$.
The set $\bar\omega_{\mathcal C,F}(c,U)$ consists of the $(\mathbb N \sqcup \{\infty\})$-indexed descending sequences of semidecidable subsets of $U$.
Note that there is no continuity condition at infinity, the last subset need only be \emph{contained} in the intersection of the earlier ones.

\subsection{Summing concrete sites}
\label{sec:sum-sites}
The fully abstract model depends on combining many different sites together.
Here we describe this process as an elementary construction for `summing' a small collection of concrete sites.

\begin{definition}\label{def:sum-site}
  Let $\{(\mathbb C_i,J_i)\}_{i \in I}$ be a (non-empty) family of concrete sites.
  Then $\sum_i \mathbb C_i$ is the category whose objects are $\coprod \mathsf{ob}(\mathbb C_i)/\sim$, where $\sim$ identifies the terminal objects in each category, and whose morphisms $(c \in \mathbb C_i) \to (d \in \mathbb C_j)$ are those functions $|c| \to |d|$ which are in $\mathbb C_i$ if $i = j$ and all constant functions if $i \neq j$.
  The coverage $\sum_i J_i$ has precisely the covers of $J_i$ for $c \in \mathbb C_i$ and $\star$ covered by the identity.
\end{definition}
It is straightforward to see that $(\sum_{i \in I} \mathbb C_i,\sum_{i \in I} J_i)$ is a concrete site.
It satisfies axioms (M) and (L) if all the $(\mathbb C_i,J_i)$ do, but it need not be subcanonical even when the $(\mathbb C_i,J_i)$ are.
Let us write $\incl j$ for the inclusion $\mathbb C_j \to \sum_{i \in I} \mathbb C_i$.
Recall that there is an adjoint triple $(\incl j)_! \dashv (\incl j)^* \dashv (\incl j)_*$ where $(\incl j)^* : [(\sum_i \mathbb C_i)^\op,\sets] \to [\mathbb C_i^\op,\sets]$ is given by precomposition with $\incl j$ and its adjoints are given by left and right Kan extension.

\begin{lemma}\label{prop:local-geometric-morphism-summed-site}
  $(\incl j)^*$ and $(\incl j)_*$ preserve sheaves and the latter is full and faithful; $(\incl j)_!$ preserves finite limits.
  A presheaf $F \in [(\sum_i\mathbb C_i)^\op,\sets]$ is a $(\sum_i J_i)$-sheaf iff $(\incl j)^*F \in [\mathbb C_j^\op,\sets]$ is a $J_j$-sheaf for all $j \in I$.
  Similarly $F$ is a concrete presheaf iff every $(\incl j)^*F$ is a concrete presheaf.
\end{lemma}
In summary, $\incl j$ induces a \emph{local geometric morphism} $\sheaves(\sum_{i \in I} \mathbb C_i,\sum_{i \in I} J_i) \to \sheaves(\mathbb C_j,J_j)$ meaning there is an adjoint triple, which we also write as $(\incl j)_! \dashv (\incl j)^* \dashv (\incl j)_*$, where $(\incl j)^*$ is precomposition with $\incl j$ and $(\incl j)_!$ is given by left Kan extension along $\incl j$ followed by sheafification, such that $(\incl j)_!$ preserves finite limits and both $(\incl j)_!$ and $(\incl j)_*$ are full and faithful.
Moreover, each $(\incl j)_!$ preserves the respective sheafified representables, and being a (concrete) sheaf for the summed site can be detected by checking under $(\incl j)^*$ for every $j \in I$.
The functors $(\incl j)^*$ are jointly faithful and satisfy $|(\incl j)^*Y| \cong |Y|$.
If $Y$ is a concrete presheaf then a function $f : |X| \to |Y|$ gives a natural transformation $X \to Y$ iff it gives a natural transformation $(\incl j)^*X \to (\incl j)^*Y$ for every $j \in I$.

We also make the following straightforward observations about monad-lifting.
\begin{proposition}\label{prop:sum-monad-lifting}
  Let $\mathbb T$ be a (strong) monad on $\sets$, and suppose that, for $i \in I$, $\mathbb T_i$ is a strong monad on $\sheaves(\mathbb C_i,J_i)$ which lifts $\mathbb T$ through the global sections function $\sheaves(\mathbb C_i,J_i) \to \sets$.
  \begin{enumerate}
  \item There is a unique strong monad $\widehat{\mathbb T}$ on $\sheaves(\sum_i \mathbb C_i,\sum_i J_i)$ which lifts each of the $\mathbb T_j$ through $(\incl j)^* : \sheaves(\sum_i \mathbb C_i,\sum_i J_i) \to \sheaves(\mathbb C_j,J_j)$.
  \item If each $\mathbb T_i$ is the partial map classifier for a generic semidecidable subobject $\Delta_i$ in $\sheaves(\mathbb C_i,J_i)$, then there is a generic semidecidable subobject $\Delta$ on $\sheaves(\sum_i \mathbb C_i,J_i)$ whose partial map classifier is $\widehat{\mathbb T}$.
  \item The colimit $\omega_{\mathbb T}$ and limit $\bar\omega_{\mathbb T}$ are sent to $\omega_{\mathbb T_j}$ and $\bar\omega_{\mathbb T_j}$ by $(\incl j)^*$.
  \end{enumerate}
\end{proposition}

\section{A fully abstract model of \texorpdfstring{\pcfv}{PCFv}}\label{sec:fully-abstract-model}

Let $I_\pcfv$ be the set of all concrete sites of the form $(\mathcal I_{\mathcal C,F},\mathcal J_{\mathcal C,F})$ where $\mathcal C$ has countably many morphisms, together with the site $\mathbb V$ (as a concrete site with trivial coverage).
For convenience, we continue to write $I_\pcfv = \{(\mathbb C_i,J_i) : i \in I_\pcfv\}$, and we write $\Delta_i$ for the specified generic semidecidable subobject in $\sheaves(\mathbb C_i,J_i)$, and $L_i$ for its associated lifting monad.
\begin{definition}
  Let $(\mathcal I,\mathcal J)$ be the sum of $I_\pcfv$ (\Cref{def:sum-site}) and let $\mathcal G = \sheaves(\mathcal I,\mathcal J)$.
\end{definition}
For $j \in I_\pcfv$, we continue to write $(\incl j)_! \dashv (\incl j)^* \dashv (\incl j)_*$ for the adjoint triple induced by $\incl j : \mathbb C_j \hookrightarrow \sum_{i \in I_\pcfv} \mathbb C_i$, as in \Cref{prop:local-geometric-morphism-summed-site}.
We write $y$ for all sheafified Yoneda embeddings.

We now show that $\mathcal G$ is a normal model of \pcfv, and subsequently a fully abstract model.
The generic semidecidable subobject $\Deltag$ is given by $\Deltag(c) = \Delta_i(c)$ for $c \in \mathbb C_i$, as in \Cref{prop:sum-monad-lifting}.
Thus the lifting monad $\liftg$ is determined by $(\incl j)^*(\liftg A) \cong L_j((\incl j)^*A)$.
We can describe $N_\mathcal{G}=\sum_0^\infty 1$ explicitly: its set of points is $N_{\mathcal{G}}(\star) = \mathbb{N}$; $N_\mathcal{G}(V)$ has only constant sequences in $\mathbb N$; and $N_{\mathcal{G}}(c,U)_{\mathcal{C},F} = \{h : U \rightarrow \mathbb{N} \mid \{h^{-1}(k) \mid k\in \mathbb{N}\}\in S^w \}$. We have the following (see \Cref{app:full:abs} for a proof):

\begin{restatable}{proposition}{liftgNatComplete}
  $\liftg(N_\mathcal{G})$ is $\liftg$-complete.
\end{restatable}

Thus $\mathcal G$ satisfies the conditions of \Cref{def:model:fgcbv}, so $\mathcal G$ admits an interpretation of \pcfv.
Moreover, it is straightforward to check that $\liftg$ preserves concrete sheaves and hence by the discussion in \Cref{sub-sec:concrete-sites} the interpretation $\den\sigma$ of each \pcfv-type $\sigma$ is a concrete sheaf.
The statement that the interpretation of \pcfv\ in $(\mathcal G,\Deltag)$  is \emph{adequate} is the same as \Cref{thm:adequacy-vsets}, and the proof is also essentially the same (see also \cite{simpson-computational-adequacy-in-an-elementary-topos}).

\subsection{Partial types}

As discussed in the introduction, our strategy for obtaining a fully abstract model is to find a model where sufficiently many morphisms are definable.
We cannot expect all morphisms to be definable since there are only countably many programs but in a normal model the interpretation of $\nat \rightarrow \nat$ always has uncountably many points.
Following \cite{milner-fully-abstract-models-of-typed-lambda-calculi} we show definability only for `partial types' --- these are finite approximations to the set of points of each type.
As discussed in \Cref{sec:kripke:rel}, the site of our sheaf model contains  `predictions' of the extent of each partial type and the system of definable functions between them. In the proof of full abstraction we will choose the prediction which is actually realized.

We do not need to consider an intrinsic definition of compactness in a normal model of \pcfv, we simply use definable idempotents to define the partial types.
The following was adapted to call-by-value from the call-by-name formulations found in \cite{DBLP:journals/iandc/OHearnR95,DBLP:books/daglib/0018087}.

\begin{definition}
  For each type $\sigma$ and $n \in \mathbb N$, define a computation $x : \sigma \cterm \psi^\sigma_n : \sigma$ by recursion on $\sigma$ where $\psi^{\nat}_n$ is ``if $x \leq n$ then $x$ else diverge'',  and $\psi^\tzero_n = x$, $\psi^\tone_n = x$,
  \begin{align*}
    \psi^{\sigma\to\tau}_n & = \ret{\lbd{u}{\letin{v}{\psi^\sigma_n[u/x]}{\letin{w}{x\ v}{\psi^\tau_n[w/x]}}}}, \\
    \psi^{\sigma + \tau}_n & = \casesum{x}{y}{\psi^\sigma_n[y/x]}{z}{\psi^\tau_n[z/x]}, \\
    \psi^{\sigma \times \tau}_n & = \letin{y}{\pi_1x}{\letin{z}{\pi_2x}{\letin{y'}{\psi^\sigma_n[y/x]}{\letin{z'}{\psi^\tau_n[z/x]}{\ret{(y',z')}}}}}.
  \end{align*}
\end{definition}

We write $h^\sigma_n : \den\sigma \to \liftg\den\sigma$ for the denotation of $\psi^\sigma_n$ in $\mathcal G$.
We will say that $h^\sigma_n$ \emph{fixes} $x \in |\den\sigma|$ if $h^\sigma(x) = \eta_{\den\sigma}(x)$.

\begin{restatable}{proposition}{idempotents}
  Each $h^\sigma_n$ is an idempotent Kleisli arrow and fixes finitely many points.
\end{restatable}
\begin{proof}[Proof notes]
  By induction on $\sigma$.
  It is clear $h^{\nat}_n$ is idempotent and fixes precisely the subobject $1 + \ldots + 1$ of $\den{\nat}$ given by the first $n+1$ points.
  For function types, $h^{\sigma\to\tau}_n$ acts on morphisms $f : \den\sigma \to \liftg\den\tau$ by $f \mapsto (h^\tau_n)^\dagger \circ f^\dagger \circ h^\sigma_n$
  (where $(-)^\dagger$ is Kleisli extension).
  The other cases are similar.
\end{proof}

From the above it is clear that we can inductively define a system of subobjects $\den\sigma_n \rightarrowtail \den\sigma$, each with only finitely many points, such that the composite $\den\sigma_n \rightarrowtail \den\sigma \to \liftg\den\sigma$ admits a retraction in the Kleisli category, making $\den\sigma_n$ a splitting of the idempotent $h^\sigma_n$.
By construction, these objects satisfy $\den{\sigma\to\tau}_n \cong \den\sigma_n \Rightarrow \liftg\den\tau_n$, $\den{\sigma+\tau}_n \cong \den\sigma_n + \den\tau_n$, and $\den{\sigma\times\tau}_n \cong \den\sigma_n\times\den\tau_n$.
Treating contexts just as product types in the obvious way, we can think of the partial types $\den\sigma_n$ as giving a `truncated' interpretation of \pcfv-types.
A computation $\Gamma \cterm t : \sigma$ denotes the morphism $\den\Gamma_n \to \liftg\den\sigma_n$ given by sequencing $\den t : \den\Gamma \to \liftg\den\sigma$ with the appropriate section and retraction.

The next lemma tells us that every type $\sigma$ is the `supremum' of a chain of partial types.
If we choose a point $x$ of $\den\sigma$, $h^\sigma(n,x)=h_n^\sigma(x)$ is its level $n$ approximation.
The existence of $h^\sigma$ means these approximations form a chain, and $H^\sigma$ witnesses that the supremum is $x$.

\begin{restatable}{lemma}{typesChainApprox}\label{lem:types:chains:of:approx}
  The assignment $h^\sigma(n,x) = h^\sigma_n(x)$ defines a morphism $h^\sigma : \omegag \times \den\sigma \to \liftg \den\sigma$ in $\mathcal{G}$, whose unique extension $H^\sigma : \baromegag \times \den\sigma \to \liftg \den\sigma$ satisfies $H^\sigma(\infty,x) = x$.
\end{restatable}
\begin{proof}[Proof notes]
  The first claim uses the fact that all types are interpreted as concrete sheaves.
  The second claim is proved by induction on $\sigma$.
  For example, when $\sigma = \nat$, $H^{\nat}(-,n)$ is eventually constant with value $n$.
  For $\sigma \to \tau$, $H^{\sigma \to \tau}(-,f)$ is the sequence with $n \mapsto H^\tau(n,-)^\dagger \circ f^\dagger \circ H^\sigma(n,-)$ (where $(-)^\dagger$ is Kleisli extension).
  For each $x \in |\llbracket \sigma \rrbracket|$, this is the diagonal of the square array $m,n \mapsto H^\tau(m)^\dagger(f^\dagger(H^\sigma(n,x)))$, so one can take the limit separately in the two indices.
\end{proof}

\subsection{Definability for partial types in \texorpdfstring{$\mathcal G$}{G}}

We show that, for each $n$, one of the sites used to obtain $\mathcal G$ was a correct prediction, and so our summed site already contains the truncated interpretation of \pcfv-types.
Let $\mathcal C_n$ be the category whose objects are types $\sigma$ and whose morphisms $\sigma \to \tau$ are morphisms $\den\sigma_n \to L\den\tau_n$ which arise as the interpretation of a term $x : \sigma \cterm t : \tau$.
Let $F_n : \mathcal C_n \to \ssppar$ map $\sigma$ to $(|\den\sigma_n|,S^{\sigma,n})$ where $P \in S^{\sigma,n}$ iff $P$ is the collection of non-empty fibres of a map $\den\sigma_n \to L\den\nat$ which arises as the interpretation of a term $x : \sigma \vdash t : \nat$.
We treat contexts $\Gamma$ as objects of $\mathcal C_n$ by identifying them with a product type.

Although the global elements of the sheafified representable $y((\sigma,U)_{\mathcal C_n,F_n})$ are naturally identified with $U$, it is not clear that there is a morphism $y((\sigma,U)_{\mathcal C_n,F_n}) \to \den\sigma_n$ corresponding to the inclusion.
Nevertheless, since the latter is a concrete sheaf there is an identification of $\den\sigma_n((\Gamma,U)_{\mathcal C_n,F_n})$ with a subset of the functions $U \to |\den\sigma_n|$.
Moreover, $\liftg(\den\sigma_n)((\Gamma,U)_{\mathcal C_n,F_n})$ can be identified with a subset of the partial functions $U \partialmap |\den\sigma_n|$, whose domain $U' \subseteq U$ is an element of $\Deltag((\Gamma,U))$, i.e.\ definable by a computation $\Gamma \cterm t : \tone$.

For convenience, let us write $\incl n : \mathcal I_{\mathcal C_n,F_n} \hookrightarrow \mathcal I$ for the inclusion of sites.
Recall from \Cref{prop:local-geometric-morphism-summed-site} that $\incl n$ induces an adjoint triple $(\incl n)_! \dashv (\incl n)^* \dashv (\incl n)_*$, where $(\incl n)_!$ preserves finite limits and representables, and $(\incl n)_!,(\incl n)_*$ are full and faithful.
The next lemma is crucial and is proved in \Cref{app:full:abs}.
Note that, in particular, it implies that every point of $\den\sigma_n$ is the interpretation of a closed value.

\begin{restatable}{lemma}{FullDefinabilityLemma}
  There is an isomorphism $y(\sigma,|\den\sigma_n|) \to (\incl n)^*\den\sigma_n$ in $\sheaves(\mathcal I_{\mathcal C_n,F_n},\mathcal J_{\mathcal C_n,F_n})$.
\end{restatable}

\begin{theorem}[Full abstraction]\label{thm:full-abstraction}
  If two \pcfv\ computations $\Gamma \cterm t,t' : \sigma$ are contextually equivalent then $\den t = \den{t'}$, and similarly for values.
\end{theorem}
\begin{proof}[Proof notes]
  The computations $t$, $t'$ denote morphisms $\den\Gamma \to \liftg \den\sigma $.
  By an induction on $\sigma$, $\den t$ and $\den {t'}$ agree on their restrictions to ${\den\Gamma}_n$: for the function type $\sigma \to \tau$ ones uses the fact that every point of ${\den\sigma}_n$ is definable and applies the induction hypothesis for $\tau$.
  It follows that ${\den t}^\dagger \circ H^\Gamma$ and ${\den{t'}}^\dagger \circ H^\Gamma$ agree on $\omegag \times \den\Gamma$ (where $(-)^\dagger$ is Kleisli extension).
  But $\liftg\den\sigma$ is $\liftg$-complete, so they also agree on $\baromegag \times \den\Gamma$.
  Evaluating at $\infty$, we get $\den t = \den {t'}$ by \Cref{lem:types:chains:of:approx}.
  The proof for values is similar.
\end{proof}

\section{Related work and research directions}
\label{sec:related-work}

\subsection{Comparison with the model of Riecke-Sandholm}

Our fully abstract model of \pcfv, $\mathcal G$, is heavily inspired by the fully abstract model for call-by-value FPC of Riecke and Sandholm~\cite{DBLP:journals/iandc/RieckeS02a}, itself inspired by~\cite{DBLP:journals/iandc/OHearnR95,sieber-1992} (see also subsequent work \cite{Mar00,MARZ2000133,DBLP:books/daglib/0018087,KK19}). Our sites $\mathcal{I}_{\mathcal C, F}$ (\Cref{def:icf}) are close to the `varying arities' of~\cite{DBLP:journals/iandc/RieckeS02a}; their index category $\mathcal C$~\cite[\S3.4]{DBLP:journals/iandc/RieckeS02a} corresponds to our $\mathcal C$, and their `path theory' $S^w$ corresponds to our \ssp\ structure $S^{F(w)}$.

The objects of our~$\mathcal G$ are in particular $\mathbb V$-sets,
and if we insist that they are moreover $\omega$-cpo's then the Kleisli category of $L$ is almost equivalent to the category $\mathcal{RCPO}$ of~\cite{DBLP:journals/iandc/RieckeS02a}. Our sheaf condition corresponds to the structure of a `computational relation' from~\cite{DBLP:journals/iandc/RieckeS02a}.

 There are some technical differences: they use directed cpo's rather than $\omega$-cpo's, and they did not require morphisms $f:v\rightarrow w \in \mathcal C$ to pull back a partition from $S^w$ to a partition from $S^v$.
But at a higher level, while it is possible that Riecke and Sandholm had sheaves and monads in mind, those concepts which are central to this paper are not explicit in~\cite{DBLP:journals/iandc/RieckeS02a}.

\subsection{Comparison with work on `Synthetic Domain Theory'}
\label{sec:sdt-review}
The vision of synthetic domain theory (SDT) is that, by working in an intuitionistic set theory, we can interpret types as sets and assume that all functions are suitably continuous. 
Our work intersects with many of the methods of this theory, even if our motivation is less philosophical and rather to use sheaf categories to build and relate models.
We comment on several aspects of SDT.
\begin{description}
\item[Partiality.]
  Our treatment of partial maps (\Cref{sec:partial-maps}) is based on~\cite{rosolini-phd} and our development of lifting monads on~\cite{mulry-partial-map-classifiers-and-partial-cccs,mulry-monads-and-algebras-in-the-semantics-of-partial-data-types}.
  In recent years the \emph{restriction categories} of \cite{cockett-lack-restriction-categories-i-categories-of-partial-maps,cockett-lack-restriction-categories-ii-partial-map-classification}
   have become increasingly popular, although these can be related to earlier methods. 
  Our construction of $\mathcal I_{\mathcal C,F}$ is reminiscent of the `splitting' of a restriction category, and our construction of $\sheaves(\mathcal I_{\mathcal C,F},\mathcal J_{\mathcal C,F})$ is reminiscent of the free cocompletion of \cite{lin-presheaves-over-a-join-restriction-category,garner-lin-cocompletion-of-restriction-categories}.
\item[Recursion.]
  Our treatment of recursion~(\Cref{sec:gener-sett-recurs}) perhaps originates in~\cite[\S5]{longley-simpson-sdt-real} or~\cite{fpp-cuboidal}; more abstract treatments were given later~\cite{oosten-simpson-sdt,reus-streicher-general-sdt}. Orthogonality also plays a central role in the representation theorem of~\cite{fiore-adt-unpublished}. SDT permits an alternative, more refined analysis of recursion, based on `replete objects'~\cite{hyland-sdt,taylor-fixedpoints-sdt}, which we have not yet pursued.
\item[Sheaf categories.]
  Much work on SDT has focused on realizability categories, but there has been substantial work on sheaf models too, beginning from Scott~\cite{scott-relating}. The idea of a Scott topos is to take sheaves on a model of the untyped $\lambda$-calculus; this is further developed in \cite[\S7.2]{rosolini-phd} and \cite[\S5]{taylor-fixedpoints-sdt}. Later work considered the monoid~$\mathbb V$ \cite{fiore-rosolini-h,fiore-rosolini-2sdt} and a stable version of it, and then general models of axiomatic domain theory~\cite{fp-adt-sdt}. But given the relevance of sheaf constructions to definability in terminating, typed calculi~\cite{DBLP:conf/tlca/FioreS99,jung-tiuryn-1993}, it is perhaps surprising that further sheaf models of SDT have not been developed. Going beyond this article, one point is that sheaf categories arguably cannot support a small complete category, which is useful for impredicative polymorphism~\cite[Ax.~3]{simpson-rosolini-lily}, although there are sheaf models of System~F nonetheless~\cite[Thm.~4.6]{pitts-polymorphism}. 
\end{description}
\subsection{Summary and outlook}
We have given a sheaf theoretic model of a call-by-value PCF (\Cref{sec:high-order-lang}) which is fully abstract~(\Cref{sec:fully-abstract-model}). Our model uses a categorical framework for partiality (\Cref{sec:partial-maps}) and recursion (\Cref{sec:gener-sett-recurs}), and is based on combining sites for sequentiality (\Cref{sec:kripke:rel})
with a site for recursion~(\Cref{sec:presheaves-vert-nat}).
The way that sites for sheaves can be combined and compared plays a crucial role.
Looking beyond this work, we anticipate that in the future it will be informative to use the flexibility of sheaves and sites to compare and combine the methods for recursion here with recent sheaf methods for other aspects of programming (e.g.~\cite{qbs,huot-staton-vakar,lmz-quantum,sherman-michel-carbin}).

\subparagraph*{Acknowledgements.} One personal starting point was recent work on Kripke logical relations models
for full abstraction in languages with effects but without recursion~\cite{KK19,kammar-katsumata-saville}. We are grateful to the authors for discussions, although we have to leave combining recursion with other effects for future work.
We thank Marcelo Fiore, Mathieu Huot, Hugo Paquet, Philip Saville, Thomas Streicher, and the anonymous reviewers for helpful feedback.

\bibliography{submitted-ssp}

\appendix
\appendixpage

\section{Proofs of technical results}

\subsection{Fixed Points}\label{app:fix}

  Let $X \in \mathbb C$ be such that $LX$ is a $L$-complete object.
  Then, for any map $f : \Gamma \times LX \to LX$, we can construct a map $\xi_f : \Gamma \to LX$ with $f(\rho,\xi_f(\rho)) = \xi_f(\rho)$ as follows.

  First define a family of maps $\mathsf{ap}_n : \Gamma \times L^n1 \to LX$ as follows: $\mathsf{ap}_0(\rho,*)  = \bot_X$ and $\mathsf{ap}_{n+1}$ the following composite:
\begin{equation*}
  \Gamma \times L^{n+1}1
  \xrightarrow{\sigma_{\Gamma,L^n1}}
  L(\Gamma \times L^n 1)
  \xrightarrow{L(\pi_1,\mathsf{ap}_n)}
  L(\Gamma \times LX)
  \xrightarrow{L(f)}
  LLX
  \xrightarrow{\mu_X}
  LX.
\end{equation*}

  The sequence $(\mathsf{ap}_n)$ defines a map $\mathsf{ap}_\omega : \Gamma \times \omega \to LX$ because it forms a cocone for diagram~\ref{eq:colimit:diag}. For this we can show by induction on $n$ that $\mathsf{ap}_{n+1} \circ (\mathsf{id}_\Gamma \times L^n(\bot_1)) = \mathsf{ap_n}$.

  Next we show that   
  \begin{equation*}
    \mathsf{ap}_\omega \circ (\mathsf{id}_\Gamma \times \mathsf{succ}_\omega) = f \circ (\pi_1 , \mathsf{ap}_\omega).
  \end{equation*}
  The sequence of maps $\mathsf{ap}_{n+1} \circ (\mathsf{id}_\Gamma \times \eta_{L^n 1})$ forms a cocone with apex $LX$ for diagram~\ref{eq:colimit:diag}, whose comparison arrow is $\mathsf{ap}_\omega \circ (\mathsf{id}_\Gamma \times \mathsf{succ}_\omega)$. Similarly the sequence $f \circ (\pi_1,\mathsf{ap}_n)$ forms a cocone with comparison arrow $f \circ (\pi_1 , \mathsf{ap}_\omega)$. So it suffices to show $\mathsf{ap}_{n+1} \circ (\mathsf{id}_\Gamma \times \eta_{L^n 1}) = f \circ (\pi_1,\mathsf{ap}_n)$ which is not hard.

  Let $\mathsf{ap}_{\bar\omega} : \Gamma \times \bar\omega \to LX$ be the unique extension of $\mathsf{ap}_\omega$.
  Observe that $\mathsf{ap}_{\bar\omega} \circ (\mathsf{id}_\Gamma \times \mathsf{succ}_{\bar\omega}) = f \circ (\pi_1 , \mathsf{ap}_{\bar\omega})$ as well.
  Then let $\xi_f(\rho) = \mathsf{ap}_{\bar\omega}(\rho,\infty)$, and now
  \begin{equation*}
    \xi_f(\rho) = \mathsf{ap}_{\bar\omega}(\rho,\infty)
           = \mathsf{ap}_{\bar\omega}(\rho,\mathsf{succ}_{\bar\omega}(\infty))
           = f(\rho,\mathsf{ap}_{\bar\omega}(\rho,\infty)) 
           = f(\rho,\xi_f(\rho))
  \end{equation*}
  as required.

Assume, as in \Cref{thm:fixed-points-x}, that $X\in\mathbb{C}$ is an $L$-algebra and $LX$ an $L$-complete object. Consider a map $g:\Gamma\times X \rightarrow X$. We will construct a fixed point $\phi_g:\Gamma\rightarrow X$ for $g$.

Using the algebra structure of $X$, $(X,\alpha)$, we can construct a map:
  \begin{equation*}
    \Gamma \times LX \xrightarrow{1\times \alpha}  \Gamma \times X \xrightarrow{g} X \xrightarrow{\eta} LX.
  \end{equation*}
  Then we can use the result from the previous paragraph to get a fixed point $\xi : \Gamma \rightarrow LX$ of this map. So the candidate fixed point for $g$ will be $\phi_g=\alpha \circ \xi$. And indeed:
  \begin{align*}
    g\circ(1,\alpha\circ\xi) &= \alpha \circ \big(\eta \circ g \circ (1\times\alpha)\big) \circ (1,\xi) &\text{because }\alpha \text{ is an algebra} \\
    &= \alpha \circ \xi & \text{because }\xi \text{ is a fixed point}.
  \end{align*} 

  Assume, as in \Cref{cor:cbv:recursion}, that $L(LB^A)$ is an $L$-complete object and $M:\Gamma\times LB^A \times A \rightarrow LB$ is a morphism. To construct a fixed point $\mathsf{rec}_M:\Gamma\rightarrow LB^A$ for $M$, notice that $LB^A$ is an algebra for $L$ because $L$ is strong, so we have:
  \begin{equation*}
    L(LB^A) \times A \xrightarrow{\sigma_{A,LB^A}} L(LB^A\times A) \xrightarrow{L\mathsf{ev}} LLB \xrightarrow{\mu_B} LB.
  \end{equation*}
  We can curry $M$ to get $\Gamma \times LB^A\rightarrow LB^A$ and then construct $\mathsf{rec}_M$ as in the previous paragraph.

\subsection{Adequacy for \texorpdfstring{$\vsets$}{vSet}}\label{app:adequacy}

\adequacy*
\begin{proof}[Proof sketch]
  Soundness is proved easily by induction on the definition of $\Downarrow_\tau$.

  Adequacy is proved using the standard method for cpo's. We define a logical relation by induction on types that says when a term is approximated by an element of the model: $\logval{\tau} \subseteq \den\tau \times \mathsf{Val}_\tau$ and $\logcomp{\tau} \subseteq \liftvsets\den\tau \times \mathsf{Comp}_\tau$. For example:
  \begin{align*}
    \logval{\tau\rightarrow\tau'} &= \{ (d,v) \mid \forall a \in \den\tau,\ w \in \mathsf{Val}_\tau.\ a \logval{\tau} w \implies (d\ a) \logcomp{\tau'} (v\ w)\} \\
    \logcomp{\tau} &= \{ (d,t) \mid \text{if } d=\eta_{\den\tau}\circ d' \text{ then } \exists w.\ t\Downarrow_\tau w \text{ and } d'\logval{\tau} w \}.
  \end{align*}
  Then we prove the fundamental property of this logical relation and show it is enough to obtain adequacy.

  The fundamental property is proved by induction on terms. For the $\mathsf{rec}$ case we prove by induction on types that all subobjects of the form $\{(-) \logcomp{\tau''} t''\}$ are closed under sups of chains. (Here a chain is a map $\omega\rightarrow \liftvsets\den{\tau''}$, and a chain with a lub is $\bar\omega\rightarrow \liftvsets\den{\tau''}$.) This replaces the proof from cpo's that the logical relation is an admissible subset.
\end{proof} 

\subsection{A fully abstract model of \texorpdfstring{\pcfv}{PCFv}}\label{app:full:abs}

\liftgNatComplete*
\begin{proof}[Proof]
  Consider an extension problem $f : (\incl j)_!y(c) \times \omegag \to \liftg(N_{\mathcal G}) $, where $c \in \mathbb C_j$, and consider two cases for $j$.
  Firstly, if $j$ is $\mathbb V$, then $(\incl j)^*f : y(c) \times \omega_\vsets \to \liftvsets(N_\vsets)$ has a unique extension to a map $y(c) \times \bar\omega_\vsets \to \liftvsets(N_\vsets)$ in $\vsets$, where the underlying function on points $\phi : |c| \times |\bar\omega_\vsets| \to |\mathbb N \cup \{\bot\} |$ is given by taking $\phi(x,\infty)$ to be the eventual value of $\phi(x,n)$ as $n \to \infty$.
  It remains to check that $\phi$ underlies a natural transformation $(\incl j)_!y(c) \times \baromegag \to \liftg(N_{\mathcal G})$ in the sheaf category $\mathcal G$.
  This is so since if $d \in \mathbb C_k$ with $k \neq j$ then $|d|$ is finite and thus for any pair $(g,h) \in ((\incl j)_!(y(c)) \times \baromegag)(d)$ we have $(\phi \circ (g,h))(y) = \phi(g(y),\min\{N,h(y)\}) = f(g(y),\min\{N,h(y)\}) \in \liftvsets(N_\vsets)(d)$ for some $N \in \mathbb N$ not depending on $y \in |d|$.
  Secondly, if $j$ is of the form $(\mathcal I_{\mathcal C,F},\mathcal J_{\mathcal C,F})$ for some faithful functor $F : \mathcal C \to \ssppar$, then since $|c|$ is finite $f$ factorizes as a retraction $(\incl j)_!y(c) \times \omegag \twoheadrightarrow (\incl j)_!y(c) \times \liftg^n 1$ for some $n$ followed by a map $(\incl j)_!y(c) \times \liftg^n 1 \to \Deltag$.
  This gives one possible extension of $f$ to $(\incl j)_!y(c) \times \baromegag$.
  Since it must also be a morphism of the underlying v-sets, it is unique.
\end{proof}

We will need the following result on preservation of exponentials, which can be extracted from the proof of Lemma A1.5.8 in \cite{John02}.
\begin{proposition}[Frobenius reciprocity]
  Let $F : \mathbb C \to \mathbb D$ be a functor between cartesian closed categories with a left adjoint $L \dashv F$.
  Then $F$ preserves a given exponential $A \Rightarrow C$ iff, for all $B \in \mathbb D$, $C$ is right-orthogonal to the canonical map $L(B \times FA) \to LB \times A$.
\end{proposition}

\FullDefinabilityLemma*
\begin{proof}
  First note that $(\incl n)^*$ is faithful on maps into concrete sheaves, and while not in general full, it is bijective on global elements.
  We proceed by induction on $\sigma$.
  Since $\den\tone_n$ is a terminal object and $\den\tzero_n$ is an initial object, both are preserved by $(\incl n)^*$ so the claim there is trivial.
  Similarly, $(\incl n)^*$ preserves sums and $y : \mathcal I_{\mathcal C_n,F_n} \to \sheaves(\mathcal I_{\mathcal C_n,F_n},J_{\mathcal C_n,F_n})$ preserves sums of types, hence the base case of $\sigma = \nat$ and the inductive case $\sigma = \sigma_1 + \sigma_2$ both hold.
  In the case of the product type $\sigma = \sigma_1 \times \sigma_2$, we have first to observe that $(\sigma\times\tau,|\den{\sigma\times\tau}|)$ is actually a product in $\mathcal I_{\mathcal C_n,F_n}$ since all global elements of $\den{\sigma_1}_n$ and $\den{\sigma_2}_n$ are definable; the claim then follows since $(\incl n)^*$ preserves products.

  The interesting case is the function types, since $(\incl n)^*$ does not preserve exponentials in general, but we will show that it does preserve the exponentials $\den{\sigma\to\tau}_n \cong \den\sigma_n \Rightarrow \liftg\den\tau_n$.
  This will suffice since we now show that $y(\sigma\to\tau,|\den{\sigma\to\tau}_n|)$ is an exponential.
  Since  $(\incl n)^*$ commutes with the lifting monad, the induction hypothesis implies that $(\incl n)^*$ is full and faithful on maps $\den\sigma_n \to \liftg\den\tau_n$, and hence all points of $\den{\sigma\to\tau}_n$ are definable.
  Then, for any $(\Gamma,U) \in \mathcal I_n$,  each map $f : y(\Gamma,U) \times y(\sigma,|\den\sigma_n|) \to L_{\mathcal C_n,F_n} y(\tau,|\den\tau_n|)$ has an underlying $f_1 : y(\Gamma \times \sigma, U \times |\den\sigma_n|) \to L_{\mathcal C_n,F_n} y(\tau,|\den\tau_n|)$ given by precomposition with $y(\Gamma\times\sigma,U \times |\den\sigma_n|) \to y(\Gamma,U) \times y(\sigma,|\den\sigma_n|)$ and thus is definable.
  Moreover, every definable function does give a natural transformation $y(\Gamma,U) \times y(\sigma,|\den\sigma_n|) \to L_{\mathcal C_n,F_n} y(\tau,|\den\tau_n|)$, whence one may deduce that $y(\sigma,|\den\sigma_n|) \Rightarrow L_{\mathcal C_n,F_n} y(\tau,|\den\tau_n|) \cong y(\sigma\to\tau,|\den{\sigma\to\tau}_n|)$.

  Now we use Generalized Frobenius reciprocity to show that $(\incl n)^*(\den\sigma_n \Rightarrow \liftg\den\tau_n) \cong (\incl n)^*(\den\sigma_n) \Rightarrow (\incl n)^*(\liftg\den\tau_n)$.
  It clearly suffices to restrict attention to those `$B$' which are representables $y(\Gamma,U)$.
  Since $(\incl n)_!(y(\Gamma,U)\times y(\sigma,|\den\sigma_n|)) \to (\incl n)_!(y(\Gamma,U)) \times \den\sigma_n$ is surjective on points, we have the uniqueness part of orthogonality.
  Now, given a map $(\incl n)_!y(\Gamma,U) \times (\incl n)_!y(\sigma,|\den\sigma_n|) \to \liftg\den\tau_n$, by precomposition we get a map $(\incl n)_!y(\Gamma\times\sigma,U\times|\den\sigma_n|) \to \liftg\den\tau_n$ whence we deduce that the underlying function is definable.
  We must show that a definable function is a natural transformation $(\incl n)_!(y(\Gamma,U)) \times \den\sigma_n \to \liftg\den\tau_n$.  
  It suffices to show the same thing with an unsheafified representable: i.e.\ to consider $\mathcal I(-,(\Gamma,U)_{\mathcal C_n,F_n}) \times \den\sigma_n \to \liftg\den\tau_n$.
  On objects of $X \in \mathcal I$ not in $\mathcal I_{\mathcal C_n,F_n}$, the set $\mathcal I(X,(\Gamma,U)_{\mathcal C_n,F_n}) \times \den\sigma_n(X)$ is indeed mapped into $L_{\mathcal E}\den\tau_n(X)$ since the left factor of the latter contains only constant functions.
  On objects $(\Gamma',U') \in \mathcal I_{\mathcal C_n,F_n}$, the same reasoning applies for constant functions $(\Gamma',U') \to (\Gamma,U)$, but for non-constant functions, which are by construction definable, we use the facts that every function in $\den\sigma_n(\Gamma',U')$ is definable and that the definable functions are closed under pairing and composition.
\end{proof}

\section{Typing rules and operational semantics for \texorpdfstring{\pcfv}{PCFv}}\label{app:typing:rules}

In this section we provide the full type system and operational semantics for the \pcfv\ language. Recall the syntax of \pcfv:
\begin{align*}
  \text{Types:}\quad \tau &\Coloneqq \tzero \mid \tone \mid \nat \mid \tau + \tau \mid \tau \times \tau \mid \tau\rightarrow\tau \\
 \text{Values:}\quad  v,w &\Coloneqq x \mid \star \mid \inl{v} \mid \inr{v} \mid (v,v) \mid \vzero \mid \suc{v} 
  \mid \lbd{x}{t} \mid \rec{f}{x}{t}\\
\text{Computations:}\quad  t  &\Coloneqq \ret{v} \mid  \casesum{v}{x}{t}{y}{t'} \mid \pi_1v \mid \pi_2v \mid v\ w
  \\
  &\mid \casenat{v}{t}{x}{t'}
  \mid  \letin{x}{t}{t'}   
\end{align*}

The typing relation is the least relation closed under the following rules:
\begin{gather*}
  \inferrule{ }{\Gamma,x:\tau \vterm x :\tau} \quad
  \inferrule{ }{\Gamma \vterm \star : \tone} \quad
  \inferrule{ \Gamma \vterm v :\tau}{\Gamma \vterm \inl{v} :\tau +\tau'} \quad
  \inferrule{ \Gamma \vterm v :\tau'}{\Gamma \vterm \inr{v} :\tau +\tau'} \\
  \inferrule{\Gamma \vterm v:\tau \\ \Gamma \vterm v':\tau'}{\Gamma \vterm (v,v'):\tau\times\tau'} \quad
  \inferrule{ }{\Gamma \vterm \vzero :\nat} \quad
  \inferrule{\Gamma\vterm v: \nat}{\Gamma \vterm \suc{v}: \nat}\\
  \inferrule{\Gamma,\,x:\tau \cterm t:\tau'}{\Gamma \vterm \lbd{x}{t} :\tau\rightarrow\tau'} \quad
  \inferrule{\Gamma,\, f:\tau\rightarrow\tau',\, x:\tau \cterm t:\tau'}{\Gamma \vterm \rec{f}{x}{t} : \tau\rightarrow \tau'} \quad
  \inferrule{\Gamma \vterm v : \tzero}{\Gamma \cterm \caseempty v : \tau} \\
  \inferrule{\Gamma \vterm v :\tau}{\Gamma \cterm \ret{v} : \tau} \quad
  \inferrule{\Gamma \vterm v :\tau+\tau' \\ \Gamma,x:\tau \cterm t : \sigma \\ \Gamma,y:\tau' \cterm t':\sigma}{\Gamma \cterm \casesum{v}{x}{t}{y}{t'} :\sigma} \\
  \inferrule{\Gamma \vterm v: \tau\times\tau'}{\Gamma \cterm \pi_1v : \tau} \quad
  \inferrule{\Gamma \vterm v: \tau\times\tau'}{\Gamma \cterm \pi_2v : \tau'} \quad
  \inferrule{\Gamma \vterm v :\tau\rightarrow \tau' \\ \Gamma \vterm w :\tau}{\Gamma \cterm v\ w : \tau'} \\
  \inferrule{\Gamma \vterm v:\nat \\ \Gamma \cterm t :\tau \\ \Gamma,x:\nat \cterm t':\tau}{\Gamma \cterm \casenat{v}{t}{x}{t'} : \tau} \quad
  \inferrule{\Gamma \cterm t :\tau \\ \Gamma,x:\tau \cterm t :\tau'}{\Gamma \cterm \letin{x}{t}{t'} : \tau'}
\end{gather*}

The big-step operational semantics of \pcfv\ is a family of relations, indexed by types, between closed computations and closed values. It is the least relation closed under the rules below:
\begin{gather*}
  \inferrule{ }{\ret{v} \Downarrow_\tau v} \quad
  \inferrule{ }{\pi_1 (v,v') \Downarrow_\tau v} \quad
  \inferrule{ }{\pi_2 (v,v') \Downarrow_\tau v'} \\
  \inferrule{t[v/x] \Downarrow_\tau w}{\casesum{\inl{v}}{x}{t}{y}{t'} \Downarrow_\tau w} \quad
  \inferrule{t'[v/x] \Downarrow_\tau w}{\casesum{\inr{v}}{x}{t}{y}{t'} \Downarrow_\tau w} \\
  \inferrule{t[(\rec{f}{x}{t})/f,\,v/x] \Downarrow_{\tau} w}{(\rec{f}{x}{t})\ v \Downarrow_{\tau} w} \quad
  \inferrule{t[v/x] \Downarrow_\tau w}{(\lbd{x}{t})\ v \Downarrow_\tau w} \quad
  \inferrule{t \Downarrow_{\tau} v \\ t'[v/x] \Downarrow_\tau w}{\letin{x}{t}{t'} \Downarrow_\tau w} \\
  \inferrule{t \Downarrow_\tau w}{\casenat{\vzero}{t}{x}{t'} \Downarrow_\tau w} \\
  \inferrule{t'[v/x] \Downarrow_\tau w}{\casenat{\suc{v}}{t}{x}{t'} \Downarrow_\tau w}
\end{gather*}

\end{document}